\documentclass[3p,times]{elsarticle}

\usepackage{ecrc}

\volume{??}

\firstpage{1}

\journalname{Theoretical Computer Science}

\runauth{}

\jid{}

\jnltitlelogo{}

\usepackage{amssymb}
\usepackage{amsthm}
\usepackage{algorithm}
\usepackage{algpseudocode}

\algtext*{EndWhile}
\algtext*{EndIf}
\algtext*{EndFor}

\newtheorem{definition}{Definition}

\newtheorem{property}{Property}

\usepackage[figuresright]{rotating}
\usepackage{subfigure}

\usepackage{color,amsmath,accents,dsfont}

\newcommand{\BN}{^{[N]}}
\newcommand{\stella}[1]{\accentset{\star}{#1}}
\newcommand{\anello}[1]{\accentset{\circ}{#1}}
\providecommand{\abs}[1]{\lvert#1\rvert}

\newcommand{\Ns}{N*}

\newdefinition{ex}[thm]{Example}

\newdefinition{rem}[thm]{Remark}

\begin{document}

\begin{frontmatter}

\dochead{}

\title{Approximate analysis of biological systems by hybrid switching jump diffusion}

\author[l1]{Alessio Angius}
\author[l1]{Gianfranco Balbo}
\author[l1]{Marco Beccuti}
\author[l2]{Enrico Bibbona}
\author[l1]{Andras Horvath}
\author[l2]{Roberta Sirovich}

\address[l1]{Universit\`a di Torino,
Dipartimento di Informatica\\
\{beccuti,angius,horvath,balbo\}@di.unito.it}

\address[l2]{Universit\`a di Torino,
Dipartimento di Matematica\\
\{roberta.sirovich,enrico.bibbona\}@unito.it}

\begin{abstract}
In this paper we consider large state space continuous time Markov chains
arising in the field of systems biology.  For a class of such models,
namely, for density dependent families of Markov chains that represent the
interaction of large groups of identical objects, Kurtz has proposed two
kinds of approximations.  One is based on ordinary differential equations
and provides a deterministic approximation, while the other uses a diffusion
process with which the resulting approximation is stochastic.  The
computational cost of the deterministic approximation is significantly
lower, but the diffusion approximation retains stochasticity and is able to
reproduce relevant random features like variance, bimodality, and tail
behavior that cannot be captured by a single deterministic quantity.

In a recent paper, for particular stochastic Petri net models, we proposed
a jump diffusion approximation that aims at being applicable beyond the
limits of Kurtz's diffusion approximation in order to cover the case when
the process reaches the boundary with non-negligible probability. In this
paper we generalize the method so that it can be applied to any density
dependent Markov chains.  Other limitations of the diffusion approximation
in its original form are that it can provide inaccurate results when the
number of objects in some groups is often or constantly low and that it can
be applied only to pure density dependent Markov chains.  In order to
overcome these drawbacks, in this paper we propose to apply the
jump-diffusion approximation only to those components of the model that are
in density dependent form and are associated with high population levels.
The remaining components are treated as discrete quantities.  The resulting
process is a hybrid switching jump diffusion, i.e., a diffusion with hybrid
state space and jumps where the discrete state changes can be seen as
switches that take the diffusion from a condition to another.  We show that
the stochastic differential equations that characterize this process can be
derived automatically both from the description of the original Markov
chains or starting from a higher level description language, like
stochastic Petri nets.  The proposed approach is illustrated on three
models: one modeling the so called crazy clock reaction, one describing
viral infection kinetics and the last considering transcription regulation.
\end{abstract}


\end{frontmatter}

\section{Introduction}\label{Sec:int}


Stochastic modeling of  the dynamics of biological systems gains in
importance as more and more evidence is gathered that randomness plays an
important role in many of these phenomena \cite{TuScBu04,Wi06}.  In most cases, as
in the pioneering works of Gillespie \cite{Gillespie_1977} and
Kurtz \cite{Ku70}, the stochastic process associated with the evolution of
the biological system is a continuous time Markov chains (CTMC).  In theory, CTMCs
can be analyzed by well-established techniques \cite{St95} to characterize
both their initial transient period and their long run behavior.  In
practice however the state space of the CTMC representation of a real phenomenon is often
so large that an exact analytical treatment is not feasible.

One approach to the analysis of these models is simulation and, starting
from \cite{Gillespie_1977}, several simulation based techniques have been
proposed.  The main difficulty lies in the facts that because of the size
of the state space many simulation runs are needed to characterize the
system, and that often the interactions occur in significantly different
time scales.  Methods to overcome these difficulties were proposed
in \cite{Gi01,RaPeCaGi03,CaGiPe05}.  Approximate analytical techniques have
also been considered.  Some examples are the following.  In \cite{DaMiWo10}
the authors propose a method that dynamically limits the state space to
those states that are of non-negligible probability.  Since the number of
states can be huge even if not all states are considered,
in \cite{MaWoDiHe10,ZhWaCa10} approximate randomization methods have been
proposed.  Another natural approach is aggregation of states which can be
done either by aggregating nearby states \cite{ZhWaCa09,CiDeHiCa09} or by
exploiting the idea of flow equivalence \cite{CoCoCo11}.  Techniques that
are based on imposing a special dependency structure on the probabilities
of the states were proposed in \cite{AnHo11b,AAAHVW}.

An important alternative to the above approaches, initiated mainly by
Kurtz, is based on constructing a simpler process to approximate the
original CTMC when it models the interaction of large groups of identical
objects (which can be members of species, or populations, or proteins, or
enzymes, etc.).  A key concept in these works is the so-called {\em density
dependent} property.  For density dependent CTMCs, as it was shown
in \cite{Ku70}, it is possible to derive a set of ordinary differential
equations (ODE) that leads to a good deterministic approximation of the
CTMC when the number of interacting objects is large.  A stochastic
approximation of density dependent processes using diffusion processes,
characterized by stochastic differential equations (SDE), was proposed
instead in \cite{kurtz1976limit}.  The ODE based approximation can be
strikingly poor when the number of interacting objects is not large enough
to rule out variability and when the model involves particular random
phenomenon - characterized by bi-modal distributions and/or switching
behaviors - that are not possible to capture with a deterministic model.  In these cases the diffusion based approximation could give better approximations, although it only works up to the first visit of the boundary of the state space. A recent review
on the application of these techniques to model chemical reactions is
given in \cite{andersonKurtz}.

In \cite{ATPN14} we proposed a jump-diffusion diffusion
approximation that aims at being applicable beyond the limit of Kurtz theory.  Namely, since originally the approximation was defined only
up to the first time when it reaches a boundary, we added in the approximating model an explicit description of the behavior at the boundaries. When a components attains a boundary, indeed, it stays there for a while and then it jumps back into the interior  mimicking the behavior of the original Markov Chain.
In this way, the approach is applicable to such systems where boundaries are
reached with non-negligible probability as it happens, for example, in ecological
models where there are species that can become temporarily extinct.

The scope of this paper is to refine and extend the jump-diffusion
approximation further so that it can be applied to get a better
approximation of more general Markov Chains. The motivation of this
extension is twofold.  First, is the case when the size of a subgroup of
objects remains constantly low so that a generalized use of the diffusion
approximation to the whole system would lead to inaccurate results.
Second, is the situation in which the approximation is not applicable
because the CTMC underlying the model does not belong to a density
dependent family, even if it has a subset of components that interact in
manners which enjoy the density dependent property.  Our proposal is to
approximate such Markov chains with a hybrid process, where our
jump-diffusion approximation is applied to those components which
correspond to groups with large number of members and which interact
according to the property of density dependence. This category of
components is referred to as {\em continuous} or {\em fluid} components and
in the more usual Piecewise Deterministic Markov Processes (PDMP) approach
they would be approximated by solving a system of switching ODEs.  The
remaining components, that we call {\em discrete}, are treated according to
the mechanisms of the original CTMC.  The process resulting from our new
approach is a hybrid switching jump diffusion (HSJD) (cf. \cite{yin}) that
uses jumps to handle both the discrete components and the behavior at the
boundaries of the state space.

The above described {\em partial fluidization} is introduced starting from
CTMCs.  Yet, as models are usually defined in higher level languages, we
show that the approximate jump diffusion process can be derived starting
from stochastic Petri nets (SPN) as well.

In some simpler cases, the proposed HSJD can be analyzed analytically by
solving the Fokker-Planck partial differential equation of the process.
Two such simple illustrative examples will be proposed. When the model is
more complex, only simulation is feasible.  We will describe a simple
algorithm for the simulation and will report the results on some
models in systems biology.

In the literature several flavors of hybrid models have been proposed and
studied in the recent years.  The PDMP class, which is strongly related to
the class of HSJD processes we propose, has also been generalized in order to
include the case in which the fluid component is a diffusion process.  A
comparison between our approach and those proposed in the literature is
deferred to Section \ref{subsec:hyb}.  Petri nets with hybrid state spaces
were introduced in \cite{HoKuNiTr98} and generalized in \cite{GrSeHoBo01}.
The aim of these original proposals was both to handle systems in which the
number of objects tend to become exceedingly large and to model
intrinsically continuous quantities (like temperatures). Processes with
hybrid state space are used also as a mean to analyze models in which not
all sojourn times are exponentially distributed.  In this context the
continuous component of the state space is used to keep track of the age of
the non-exponential durations.  An important work in this direction is by
Cox \cite{PSP:2044020}.  A recent book on hybrid switching diffusion
is \cite{yin} which concentrates mainly on the mathematical theory for such
processes.

This paper is organized as follows.  Section~\ref{Sec:mat} is devoted to
provide the necessary background on density dependent families of CTMCs.
In Section~\ref{Sec:the} we derive the stochastic differential equations
that characterize the proposed hybrid switching jump diffusion
approximation.  Simple illustrative examples that can be handled
analytically are shown in Section~\ref{Sec:analytic}.  Simulation based
numerical illustration is provided in Section~\ref{Sec:res}.  Conclusions
are drawn in Section~\ref{Sec:con}.  An Appendix contains some further
material and some more details.

\section{Background}\label{Sec:mat}

This section is devoted to the necessary background on density dependent
processes.  We follow the exposition we provided in \cite{ATPN14} adding
different examples (described in \ref{sec:dd2} and in \ref{app:dd}) and
omitting some details that are not strictly necessary to the scope of this
paper.  In particular, a notable difference with respect to \cite{ATPN14}
is that we do not require that the model is with bounded state space.

\subsection{Density dependent CTMCs\label{sec:dd}}

Density dependent CTMCs often arise when a model describes the interaction
of groups of identical objects.  The term object is used intentionally to
indicate that density dependent processes are present in many contexts.
When modeling biological systems the objects are, for example, enzymes,
proteins or members of populations.  In networks of queues the objects are
customers.  Informally, the necessary condition for a model to be density
dependent is that the rate of the interactions depends on the density of
the population levels and not on the absolute population values.

Hereinafter we provide the definitions that are necessary for the rest of
the paper.  We will denote with $\mathbb{R}$, $\mathbb{Q}$, $\mathbb{Z}$
and $\mathbb{N}$ the sets of real, rational, integer, and natural numbers,
respectively. Given a positive constant, $r$, we will denote with
$\mathbb{R}^{r}$ the $r$--dimensional cartesian product of the space
$\mathbb{R}$. The letter $u$ will be used to indicate the time index
ranging continuously between $[0, +\infty)$ or $[0,T]$ when specified. The
discrete states of a continuous time Markov chain will be denoted as $k$ or
$h$ and range in the state space that is included in $\mathbb{Z}^{r}$. We
will always consider the abstract probability space to be given as
$(\Omega,\mathcal{F},\mathbb{P})$, where $\Omega$ is a non empty set,
$\mathcal{F}$ is a $\sigma$-algebra on $\Omega$ and $\mathbb{P}$ is the
probability measure. Furthermore, $\mathbb{E}$ will denote the expectation
with respect to $\mathbb{P}$.  The formal definition of a family of density
dependent CTMCs is the following \cite{Ku70}.

\begin{definition}\label{def:density_dependent}
A family of Markov chains $X^{[N]}(u)$ with parameter $N$ and with state
space $E^{[N]} \subseteq \mathbb{Z}^{r}$, is called \emph{density dependent} iff
there exists a continuous non-zero function
$f: \mathbb{R}^r \times \mathbb{Z}^{r} \rightarrow \mathbb{R}$ such that
the instantaneous transition rate (intensity) from state $k$ to state $k+l$
can be written as
\begin{equation}\label{eq:form}
q^{[N]}_{k,k+l}=N f\left(\frac{k}{N},l\right), \quad l \neq 0.
\end{equation}
\end{definition}

The indexing parameter of the family, $N$, has different meanings depending
on the context. It can be the size of the considered area, the total number
of considered objects, or the volume in which the interactions take place.
The first argument of $f$ in (\ref{eq:form}) is either the density
associated with state $k$ (if $N$ is the area or the volume of the
interactions), or the normalized state (if $N$ is the total population
size).  The second argument is the vector that describes the effect of a
transition on the state.  The consequence of the fact that every member of
the family shares the form given in (\ref{eq:form}) is twofold.  First, in
every CTMC of the family, the transitions have the same effect on the
state.  Second, given a state $k$, the intensities of the outgoing
transitions, $q^{[N]}_{k,k+l}$, depend on $k/N$ (and not on the state
itself), and are proportional to the indexing parameter $N$.  In the
following we denote the set of possible state changes by $C$, i.e.,
$C=\{l~|~l \in \mathbb{Z}^{r}, l \neq 0,
\exists k \in E^{[N]}: q^{[N]}_{k,k+l}\neq 0\}$, and the possible state
changes from a given state $k$ by $C_k$, i.e.,
$C_k=\{l~|~l \in \mathbb{Z}^{r}, l \neq 0, q^{[N]}_{k,k+l}\neq 0\}$.  An
example for a family of density dependent CTMCs is provided
in~\ref{app:ddex}.

Definition \ref{def:density_dependent} can be extended in order to include
a larger class of models that can still be treated in the same framework.
This leads to the definition of {\em {near density dependence}} where we
require that the transition rates tend to the form given in
(\ref{eq:form}), when the indexing parameter tends to infinity.

\begin{definition}\label{def:nearly-density_dependent}%
A family of Markov chains $X^{[N]}(u)$ with parameter $N$ and with state
space $E^{[N]} \subseteq \mathbb{Z}^{r}$, is called \emph{nearly density dependent}
iff there exists a
continuous non-zero function
$f: \mathbb{R}^r \times \mathbb{Z}^{r} \rightarrow \mathbb{R}$ such that
the instantaneous transition rate (intensity) from state $k$ to state $k+l$
can be written as
\begin{equation}\label{eq:qnearly}
q^{[N]}_{k,k+l}=N \left[ f\left(\frac{k}{N},l\right) + O\left( \frac{1}{N} \right) \right], \quad l \neq 0.
\end{equation}
\end{definition}

In the following subsection we provide an example of a density dependent
process.  A further example and a discussion of the properties of density
dependence are provided in \ref{app:dd}.

\subsection{Mass action chemical kinetics\label{sec:dd2}}
Chemical reaction models with rates that follow the law of mass action are
(nearly) density dependent families of Markov Chains.  As an example let us
consider the following chemical reactions
\[
A+B \stackrel{\nu_1}{\rightarrow} 2A,~~ 2A\stackrel{\nu_2}{\rightarrow} A+B, ~~2A+B \stackrel{\nu_3}{\rightarrow} A+B
\]
and assume that the intensities of the reactions follow the stochastic law of
mass action.  This means that the intensity of a reaction is proportional to
the number of distinct ways the molecules can form its input. Further, it is
inversely proportional to $V^{n-1}$ where $V$ is the volume and $n$ is the
number of molecules that form the input of the reaction.  The reason for
this is that the bigger the volume the less probable that the molecules on
the left hand side of the reaction collide.  Accordingly the intensities are
\[
q_{(i,j),(i+1,j-1)}=\nu_1 \frac{ij}{V},~~~
q_{(i,j),(i-1,j+1)}=\nu_2 \frac{i(i-1)}{2V},~~~
q_{(i,j),(i-1,j)}=\nu_3 \frac{i(i-1)j}{2V^2}
\]
where we assumed that the state is described by a pair $(i,j)$ with $i$
providing the number of molecules of $A$ and $j$ the number of molecules of
$B$.  The above intensities can be rewritten as
\[
q_{(i,j),(i+1,j-1)}=V \nu_1  \left(\frac{i}{V}\right) \left(\frac{j}{V}\right),~
q_{(i,j),(i-1,j+1)}=V\left(\frac{\nu_2}{2} \left(\frac{i}{V}\right)^2-
\frac{\nu_2}{2V}\left(\frac{i}{V}\right)\right),~
q_{(i,j),(i-1,j)}=V\left(\frac{\nu_3}{2} \left(\frac{i}{V}\right)^2\left(\frac{j}{V}\right)-
\frac{\nu_3}{2V}\left(\frac{i}{V}\right)\left(\frac{j}{V}\right)\right)
\]
where the first intensity is in exact density dependent form while the
other two contain additional terms in the order of $O(1/V)$ once $i$ and
$j$ are fixed.  It turns out that this difference does not preclude the use
of the approximation framework we consider in this paper.   For this reason
near density dependence is introduced in Definition~\ref{eq:qnearly}.

\subsection{Approximations of Density dependent CTMCs\label{sec:dd3}}

Let us turn our attention now to two approximations of density dependent
CTMCs, the limiting deterministic process introduced in \cite{Ku70} and the
sequence of diffusion processes introduced in \cite{kurtz1976limit}.  Both
of these approximations rely on processes with continuous state space and
thus fall into the category of ``fluid'' approximations.  The first
approximation employs a set of ordinary differential equations (ODEs) with
one equation per group, i.e., it provides a deterministic approximation of
the stochastic behavior of the system.  It was shown in \cite{Ku70} that,
if the number of interacting objects tends to infinity, there exists a
formal relation between the approximation provided by the ODEs and the
original process.  In practice, for a given finite number of interacting
objects, the approximation is usually seen as a mean to provide the
approximate expected number of objects for each group.  The second
approximation employs instead stochastic differential equations (SDEs) and
is referred to as the diffusion approximation of the process.  As it was
shown in \cite{kurtz1976limit}, also in case of the diffusion approximation
there exists a formal relation between the original process and the
approximation, but, in contrast to the deterministic one, this relation
holds for any number of interacting objects and not only in the limiting
case.  A crucial difference between the two approximations is that the
deterministic one provides only the approximate mean behaviour of the
random variables of interest while the diffusion approximation leads to
their approximate joint distribution.

In order to introduce the approximations we need to define the family of
normalized CTMCs given as $Z^{[N]}(u)=\frac{X^{[N]}(u)}{N}$ with state
space $S\BN$, which is also referred to as the density process.  Note that
using the normalized CTMCs the state spaces of all the members of a density
dependent family are brought to the same scale, and thus become
comparable.

Based on the properties summarized in \ref{app:prop}, the following
result was shown by Kurtz in~\cite{Ku70}.  Given a nearly density dependent
family of CTMCs $X^{[N]}(u)$, if the limit of the initial conditions tends to
$z_0$, i.e.,
\[
\lim_{N \rightarrow \infty}
Z^{[N]}(0)=\lim_{N \rightarrow \infty}\frac{X^{[N]}(0)}{N}=z_{0}
\]
and the
function
\[
F(y) = \sum_{l\in C} l f\left(y,l\right)
\]
satisfies some relatively mild conditions, then the density process
$Z^{[N]}(u)$ converges to a deterministic function $z(u)$.  The function
$z(u)$ solves the system of ODEs \footnote{Equation \eqref{eq:odeK} is
equivalent to the form $\frac{dz(u)}{du}= F(z(u))$. We have chosen the
``differential'' form written in \eqref{eq:odeK} to be consistent with the
notation that will be introduced for the stochastic differential
equations.}
\begin{equation}\label{eq:odeK}
\begin{aligned}
d z(u) &=F(z(u))du,\\
z(0)&=z_{0}.
\end{aligned}
\end{equation}
The convergence is in the following sense: for every $\delta >0$ we have
\begin{align}\label{eq:convergenceK}
\lim_{N \rightarrow \infty}
\mathbb{P}\left\{
\sup_{u\leq T}\left|
Z^{[N]}(u)-z(u)
\right|>\delta
\right\}=0.
\end{align}
where $T$ is the upper limit of the considered finite time horizon.

The function $z(u)$ is usually interpreted as the asymptotic mean.  The
difference $Z^{[N]}(u)-z(u)$ can be seen instead as the ``noisy'' part of
$Z^{[N]}(u)$.  It was shown in \cite{Ku70} that for $N \rightarrow \infty$
the density process $Z^{[N]}(u)$ flattens out at its mean value and that
the magnitude of the noise around the mean is\footnote{When used to compare
a pair of stochastic processes $A(u)$ and $B(u)$, the $O$ notation has the
following precise meaning: $A(u)-B(u)= O\big(g(N)\big)$, for some function
$g(N)$ that is infinitesimal when $N \rightarrow \infty$, if and only if
there exists an almost surely finite random variable $\Gamma_{T}$ with
finite moments of any order, such that $ \sup_{u\leq
T}\abs{A(u)-B(u)}\leq \Gamma_{T}g(N)$.}
\begin{align}
\label{eq:odediff}
Z^{[N]}(u) - z(u)= O\left( \frac{1}{\sqrt{N}} \right).
\end{align}

In practice, the convergence given in eq.~\eqref{eq:convergenceK} is often
used in case of a finite $N$ to approximate the stochastic process
$X^{[N]}(u)=\Ns Z^{[N]}(u)$ with the deterministic function $x^{[N]}(u)=\Ns z(u)$.  This
approximation disregards the noise term which is in the order
$\Ns O(1/\sqrt{N})=O(\sqrt{N})$ that is small compared to the order of the
mean (that is $N$), but not in absolute terms.  Moreover, it ignores every
detail of the probability distribution of $X^{[N]}(u)$ except for its mean.
It is easy to see that there are cases, e.g. multimodal or highly variable
distributions, where the mean gives little information about the actual
location of the probability mass, cf. \cite{JANE}.

Let us stress that the convergence holds only if
$\lim_{N \rightarrow \infty}
Z^{[N]}(0)=\lim_{N \rightarrow \infty}\frac{X^{[N]}(0)}{N}=z_{0}$, meaning
that the corresponding sequence of initial conditions $X^{[N]}(0)$ needs to
grow linearly with $N$.  This implies that if $X^{[N]}(u)$ is multivariate
then each non-zero entry of the vector describing the initial state of the
process has to grow with the same rate.

An approximation of a density dependent family $X^{[N]}$ which preserves
its stochastic nature and has a better order of convergence was proposed
in \cite{kurtz1976limit,kurtz1978strong}.  In order to introduce this
approximation, let us denote by $H(S\BN)$ the convex envelope of the state
space of the density process $Z^{[N]}$ which can be seen as the potential
state space of the continuous approximation of the discrete process.

In \cite{kurtz1976limit,kurtz1978strong} it has been shown that there
exists an open set $E \subset H(S\BN)$ (i.e., a subspace that does not
contain the boundaries of $H(S\BN)$) in which the density process $Z^{[N]}$
can be approximated by a diffusion $Y^{[N]}(u)$ with state space $E$.  The
diffusion $Y^{[N]}(u)$ is characterized by the system of SDEs
\begin{align}\label{eq:sde_ctmc}
dY^{[N]}(u)  =
F(Y^{[N]}(u)) du
 + \sum_{l \in C} \frac{l}{\sqrt{N}}  \sqrt{f(Y^{[N]}(u),l)}\;dW_{l}(u)
\end{align}
where the $\{W_{l}(u)\}$ are independent standard one-dimensional Brownian
motions and $f$ is given in eq.~\eqref{eq:form}. The approximation holds up
to the first time $Y^{[N]}(u)$ leaves $E$.

The structure of eq.~\eqref{eq:sde_ctmc} is the following: the first term
is the same that appears in eq.~\eqref{eq:odeK}, while the second term
represents the contribution of the noise and is responsible for the
stochastic nature of the approximating process $Y^{[N]}$.  A further
relation between eq.~\eqref{eq:sde_ctmc} and eq.~\eqref{eq:odeK} can be
obtained by considering that the stochastic part of the equation is
proportional to $1/\sqrt{N}$, meaning that as $N \rightarrow \infty$, this
term becomes negligible and $Y_{\infty}(u)$ solves the same ODE written in
eq.~\eqref{eq:odeK}.  Let us remark that the construction of such noise is
not based on an ad hoc assumption, but is derived formally from the
original Markov chain.

A rigorous mathematical treatment of SDEs can be found in
\cite{klebaner, rogers} where It$\bar{\mbox{o}}$ 
calculus is introduced. In the physical literature the notation
$\frac{dW(u)}{du}=\xi(u)$ is often used even if Brownian motion is nowhere
differentiable and $\xi(u)$ is called a
\emph{gaussian white noise}. This SDE approach that goes back to the
already cited \cite{kurtz1976limit,kurtz1978strong} has been applied in
many contexts, e.g., it is used under the name of \emph{Langevin equations}
to model chemical reactions in \cite{gillespie}.  Let us recall that
chemical reaction models are nearly density dependent.

As for the relation of the diffusion approximation and the original density
process, in \cite{kurtz1976limit} it has been proven that, for any finite
$N$, we have
\begin{align}
Z^{[N]}(u)- Y^{[N]}(u)= O\left( \frac{\log N}{N} \right)
\end{align}
which, compared to eq.~\eqref{eq:odediff}, is a better convergence rate.
Thus, the process $\Ns Y^{[N]}(u)$ approximates the CTMC $X^{[N]}(u)$ with
an error of order $\log N$ which is much lower than the $\sqrt{N}$ of the
deterministic fluid approximation.

Finally, let us stress that the approximation is valid only up to the first
exit time from the open set $E$.  For many applications the natural state
space is bounded and closed and the process may reach the boundary of $E$
in a finite time $\tau$ with non-negligible probability.  In such cases,
since the approximating process $Y^{[N]}(u)$ is no longer defined for any $u
\geq \tau$, the diffusion approximation is not applicable.  To overcome
this limitation suitable boundary conditions must be set and this problem,
that was considered neither in \cite{kurtz1976limit} nor in
\cite{kurtz1978strong}, will be tackled in Section~\ref{ctmc2jd}.

In \ref{sec:pn} it is shown that density dependent families of CTMCs
(including models of chemical kinetics) often arise from Stochastic Petri
Nets (SPNs) models and we describe how to translate the general theory here
proposed into the language of SPNs.

\section{Jump-diffusion approximations and the hybrid switching extention}\label{Sec:the}

In many biological systems of practical interest the applicability of the
fluid approximations we have introduced in Section~\ref{Sec:mat} is limited
by several factors: (a) the presence of boundaries in the state space that
are visited with non-negligible probability, (b) the explicit violation of
the density dependent property, (c) the low population levels that make the
fluid approximation inaccurate, and (d) the presence of ``slow components''
that renders the fluid approximation ineffective. The aim of this section
is to introduce a new model that relies on hybrid switching jump diffusion
(HSJD) processes and that is able to give a reliable approximation of
Markov models for reaction networks even in the presence of these factors.

\subsection{From CTMCs to jump diffusion processes\label{ctmc2jd}}

At the boundary of the state space, indeed, some of the state changes that
are possible in the interior are not enabled any longer (e.g. an enzymatic reaction cannot occur if there are no enzyme molecules available). If the Markov chain spends some time at the boundary, it may happen that special behaviors, which do not occur or occur
rarely in the interior, become significant and cause the appearance of
different modes in the joint probability distribution.  Such multi-modal
behavior cannot be captured by a deterministic approximation which at most
can keep track of the mean value of the marginal distributions of each
component (cf. \cite{srivastava}).  Not even the diffusion
approximation given in eq.~\eqref{eq:sde_ctmc} solves the problem since it is defined
only in the interior of the state space and when the boundary is reached
its validity ceases unless suitable boundary conditions that are not
specified in the original literature are imposed.  In \cite{ATPN14} the
problem was solved in the context of density dependent Markov models that
are described by SPNs whose transitions fire with rates proportional to their enabling degrees (i.e., depend on the number of tokens in their input places)
and whose places are all covered by P-invariants (i.e., the state space is
bounded; for more details on the SPN formalism see \cite{BOABCDF95}).  In
particular, in \cite{ATPN14} a jump diffusion approximation has been
introduced that we now recast in the more general setting of
nearly density dependent models.

Let us consider a nearly density dependent Markov chain $X \BN(u)$ and its
normalized version $Z \BN(u)$. $Z \BN$ has state space
$S\BN \subseteq \mathbb{Q}^{r}$ and it is such that the possible values of
some of its components can be bounded by a minimal and/or by a maximal
value.  Such minimal and maximal values are present, for example, in a
chemical reaction model where the number of molecules of a given chemical
compound cannot become negative, or in a population model, where any given
sub--population can neither become negative nor can exceed the total
population size.

The instantaneous transition rates of the normalized
Markov chain $Z \BN$ will be denoted by $p\BN_{x,x+l/N},
x \in \mathbb{Q}^{r}, l \in \mathbb{Z}^{r}$.  We will use the same notation
for the transition rates extended in the natural way to the state space of
the diffusion approximation where $x \in \mathbb{R}^{r}$.  Let us recall
that the natural state space of the diffusion approximation $Y \BN(u)$ is
the convex hull of the state space of the Markov chain,
$H(S\BN) \subset \mathbb{R}^{r}$.  Notice however that $Y \BN$ is defined
only in the interior of such space and up to the first time the process
visit its boundary.  In order to extend the diffusion approximation to include
the boundary of the state space, we introduce a new approximating process
$\tilde{Y}\BN(u)$ which behaves like the diffusion approximation ${Y}\BN$
in the interior of the state space and that, when the boundary is reached,
mimics the jump behavior of the original Markov chain.  We describe the
process $\tilde{Y}\BN(u)$ more formally in the following paragraphs.

In order to identify those components that are actually at their minimal or
maximal values, we define a map $B:
H(S\BN) \rightarrow \mathcal{P}(\{1,\cdots, r\})$, from the state space of
the diffusion to the power set of the set of indexes, which, given a state
$x$, provides the set of indexes of the extremal components.  Hence,
$B(\tilde{Y}^{\BN}(u))$ is the set of indexes of the components that are on
the boundary at time $u$ \footnote{Let us suppose to have a process with two components, $X_t$ and $Y_t$, both bounded between $0$ and $1$. The map $B$ applied to the vector $(X_t,Y_t)$ returns the set of the indexes of the components at the boundary. If at time $t=7$ the process is in $(X_7=0.34,Y_7=1)$, the map $B$ returns $\{2\}$, meaning that the second component, namely $Y_t$, lies on one of the boundaries.}.  Let us notice that if no components of the
actual state $\tilde{Y}^{\BN}(u)$ attain the boundaries,
then $B(\tilde{Y}^{\BN}(u))=\varnothing$ and $\tilde{Y}^{\BN}(u)$ has the
same behavior as the diffusion $Y^{\BN}(u)$.  As soon as
$B(\tilde{Y}^{\BN}(u))$ becomes non-empty, the set of possible state changes
$C_{x}=\{l \in \mathbb{Z}^{r} : l \neq 0 ,\; p_{x,x+l}\BN \neq 0\}$
is split dynamically (depending on the current state $x=\tilde{Y} ^{\BN}(u)$)
into two subsets, $\stella{C}_{x}$ and $\anello{C}_{x}$. The former
contains those state changes that either modify a component which is
extremal or which are such that the corresponding rate depends on an extremal
component, i.e.
\[ \stella{C}_{x}= \left\{ l : \exists \; i \in B(x) \text{ such that either } l_{i} \neq 0 \text{ or } \frac{\partial p \BN_{x,x+l/N}}{\partial x_{i}} \neq 0 \right\},\]
the latter is the complement set, i.e. $\anello{C}_{x}=C_{x}-\stella{C}_{x}$.


As long as the state changes included in $\stella{C}_{\tilde{Y}^{\BN}(u)}$
do not occur, the subsystem made of the components with indexes not
included in $B(\tilde{Y}^{\BN}(u))$ can still be approximated by
diffusion which are analogous to eq.~\eqref{eq:sde_ctmc} except
that the sums are restricted to the state changes in
$\anello{C}_{\tilde{Y}^{\BN}(u)}$.

The events included in $\stella{C}_{\tilde{Y}^{\BN}(u)}$ encode the behavior at the boundary. We keep them discrete and we treat them as a jump process
which is responsible for all the events of the type ``the $i$-th component
leaves the boundary''. The amplitudes and the intensities of the jumps are
formally taken from the original CTMC and depend on the entire state of the
process, i.e., on all its components, no matter whether they are at the
boundary or not.  The approximating jump diffusion $\tilde{Y} ^{\BN}(u)$
which embodies both the fluid evolution and the discrete events solves the
following system of SDEs
\begin{align}\label{eq:Jsde}
d \tilde{Y} ^{\BN}(u) = \hspace{-2mm}
\sum_{l \in \anello{C}_{\tilde{Y} ^{\BN}(u)}} \hspace{-1mm} l f \left( \tilde{Y} ^{\BN}(u),l \right) du
+ \frac{1}{\sqrt{N}} \sum_{l \in \anello{C}_{\tilde{Y} ^{\BN}(u)}} \hspace{-1mm} l \sqrt{f \left( \tilde{Y} ^{\BN}(u),l \right) }dW_{l}(u)
+ \hspace{-2mm}\sum_{l \in \stella{C}_{\tilde{Y} ^{\BN}(u)}} \hspace{-1mm}\frac{l}{N}\;  dM^{\BN}_{l}(u)
\end{align}
where  $M^{\BN}_{l}(u)$ is the counting process that describes how many events with state change $l$ occurred in the time interval $(0,u]$ and whose intensity is given by
\[
\mu_{l}(\tilde{Y} ^{\BN}(u^-))= p\BN_{\tilde{Y} ^{\BN}(u^-),\tilde{Y} ^{\BN}(u^-)+\frac{l}{N}}
\]
which depends on the actual state of the process $\tilde{Y} ^{\BN}$ right
before the jump\footnote{A function $f$ evaluated in $u^-$ is defined as
the left-sided limit $\displaystyle {f(u^-)= \lim_{x\rightarrow u^-}
f(x)}$.}. Let us recall that a counting process is a stochastic process
with positive, integer and increasing values. It is the natural model for
the number of outcomes in a system over time.  As the diffusion given
in \eqref{eq:sde_ctmc}, also \eqref{eq:Jsde} has its proper mathematical
definition in its integral form, where the integrals with respect to the
Brownian motions $W_l$ and with respect to the counting processes
$M^{\BN}_l$ have to be defined in the general It\^o theory of integration
with respect to semimartingales \cite{klebaner,rogers}.

Equation \eqref{eq:Jsde} is a system of equations, one for every component
of the process.  This might seem contradictory with our previous
description according to which only the components not included in
$B(\tilde{Y} ^{\BN}(u))$ are fluidized.  Let us however remark that the
fluid increments in the first sum of \eqref{eq:Jsde} do not affect the
component at the boundary since if $l \in \anello{C}_{x}$ then the $i$th
entry of $l$ is $0$ for any $i \in B(x)$.  On the other hand, a component
that is not at the boundary at a given time is moved by the continuous
compounding of the fluid increments that sums up with the effect of the
jumps.

\subsection{From CTMCs to hybrid switching diffusions with jumps} \label{subsec:hyb}

Many real systems violate the basic assumption that justifies the fluid
approximation, i.e., that events are very frequent and cause very small
state changes.  An example is a Markov chain in which some of the
transition rates have the nearly density dependent form of
\eqref{eq:qnearly} and others do not. It is reasonable to
``fluidize'' the system partially, making the effect of the density
dependent state changes continuous, but keeping the other state changes as
discrete events. Another example is a Markov model whose components have
values ranging in different scales.  For example, when some components of
the system model resources with limited availability.  Even if the rates
are all nearly density dependent, it is natural to consider a family of
initial conditions where the components not related to the resources grow
large with the indexing parameter of the family, while those modeling
resources are kept fixed.  Accordingly, we propose to fluidize only those
components whose increase is not in contrast with the modeling assumptions.
Moreover, if the transition rates follow the law of mass action given
in eq.~\eqref{eq:lme}, the reactions that involve reactants of which there are
only a few are typically slower than those that involve the abundant
species, introducing different time scales in the reaction rates.  A better
approximation is achieved if the components with small values and the
reactions with law rates are kept discrete, while the rest is fluidized.

Different hybrid approaches of this kind have been adopted in the
literature and we list a few here for comparison. A special mention has to
be done to Piecewise Deterministic Markov Processes (PDMPs, cf. \cite{pdmp}
for the mathematical theory and \cite{hybridLimit, pola2003stochastic,
caravagna,menz, verena,luca2010} for applications in computer science and biology)
that, as we shall see, are the ODE based counterpart to our HSJD approach.
Even hybrid approaches based on diffusion have already appeared in the
literature, mainly using switching diffusions (see \cite{yin} for the
mathematical theory and \cite{intep, haseltine, salis, pola2003stochastic}
for applications) but, as far as we know, no treatment of the behavior at
the boundaries has been included into such models. Other general and
popular approaches to hybrid systems where the fluid components are
evolving according to a diffusion have been studied
in \cite{gshs,lyg1,pola2003stochastic} under the name of (Generalized)
Stochastic Hybrid Models. Such models have a different origin (aircraft
dynamics) and have not been introduced as approximations of Markov
chains. In their dynamics, reaching of the boundaries of the state place
has a role, but it is a different one. Moreover, the partitioning of the
state space between fluid and discrete components is static, independent of
the system evolution.  The specific contribution of this paper is to adopt
the jump diffusion paradigm for the partially fluidized system that allows
the fluid components to visit the boundaries and to jump back in the
interior of the state space in almost the same way as the original chain
does.

For the sake of clarity, we will first introduce the process disregarding
the behavior at the boundary. The machinery that allows the visit of the
boundaries will be added later on.

Let us hence introduce a partial fluidization of a family of
$r$-dimentional Markov chains $X \BN$.  The formal requirements on $X \BN$
for the approximation to hold will be made clear after having introduced
the necessary notation.  Intuitively, we need that a subsystem of the
original model containing a subset of its components and a subset of its
possible state changes is nearly density dependent.  This subsystem is
fluidized and all the rest of the system (the ``problematic'' part) is
left discrete.

We assume that we are given the partition of the set of indexes $\{
1,\dots, r \}$ defined by two disjoint subsets: $F$, with cardinality $n$,
and $D$, with cardinality $m$ such that $n+m=r$, and $F \cup D = \{
1,\dots, r \}$.  The components $X \BN_{i}, i \in F,$ will be fluidized and
their initial condition will be proportional to the indexing parameter $N$.
The initial condition for the components $X \BN_{i}, i \in D,$ will remain
instead constant as $N$ increases.  The partition of the components is
supposed to be explicitly given as part of the model itself.  The process
we aim to approximate is the partially scaled Markov chain $Z \BN$ in which
only the components in $F$ are divided by $N$ (and will be approximated by
a diffusion) while the components in $D$ are kept as they are.
Consequently, $Z_{i} \BN = X_{i} \BN$ for all $i \in D$ and
$Z_{j}\BN=\frac{X_{j}\BN}{N}$ for all $j \in F$. Let us denote the state
space of $Z_{j}\BN$ by ${S'}\BN $.  A transition of the chain $X \BN$
causes a state change from state $k$ to state $k+l$ with a rate
$q_{k,k+l}\BN$. The corresponding transition of the partially scaled chain
$Z\BN$ moves the chain from state $x$ to state $x+l'$ with rate
$p\BN_{x,x+l'}=q_{k,k+l}\BN$ where the state change $l'$ is such that
$l'_{i}=l_{i}$ for all $i \in D$ and $l'_{j}=\frac{l_{j}}{N}$ for all
$j \in F$.

The partition of the components requires to partition the possible state
changes $C=\{ l \in \mathbb{Z}^{r} : \; l \neq 0 \text{ and } \exists\; x :
p\BN_{x,x+l'} \neq 0 \}$ into two disjoint sets $C^{F}$ and $C^{D}$, which
distinguish the fluid events that happen continuously by diffusion and the
discrete ones that happen by jumps.  This partitioning is made on the basis
of the following criteria.  Those transitions that modify the discrete
components necessarily cause discrete increments and hence belong to
$C^{D}$.  The events whose rates depend on a discrete component are also
treated as discrete because their rates could be too slow and the number of
involved object could be too few to justify a continuous approximation.
The remaining events, which are with rates in the nearly density dependent
form, are in $C^{F}$.  Formally we have
\[
C^{F}= \left\{ l \in C : \;   l_{i}=0\ \text{ and } \frac{\partial p_{x,x+l'}\BN}{\partial x_{i}} =0, \; \forall i \in D, x \in {S'} \BN\right\}.
\]
Let us remark that the condition $l_{i}=0, \forall i \in D,$ is strictly
necessary for preserving the discreteness of the components in $D$ while
the condition regarding the transition rates is introduced mainly to
increase the quality of the approximation and might be dropped in the case
one needs to simulate the model faster (but obviously with less
accuracy).  The remaining state changes are included in $C^{D}=C-C^{F}$. Let
us stress that the above partition of the state changes is static,
i.e., it does not depend on the current state of the process.

We approximate the partially scaled process $Z\BN$ with the process
$\Upsilon \BN$ whose components evolve with hybrid switching jump
diffusion dynamics described by the following equations:
\begin{align}
d\Upsilon \BN_{j}(u) &= \sum_{l \in C^{F}} l_{j} f\left( \Upsilon\BN (u),l \right) du
+  \sum_{l \in C^{F}} \hspace{-1mm}  \frac{l_{j}}{\sqrt{N}} \sqrt{f \left( \Upsilon \BN(u),l \right) }dW_{l}(u)
+ \hspace{-2mm}\sum_{l \in C^{D}} \hspace{-1mm}\frac{l_{j}}{N}\;  dJ^{\BN}_{l}(u), \quad \quad j \in F, \nonumber\\
d\Upsilon \BN_{i}(u) &= \sum_{l \in C^{D}} \hspace{-1mm} l_{i}\;  dJ^{\BN}_{l}(u), \quad \quad i \in D
\end{align}
where $J\BN_{l}(u)$ is the process counting the events that causes a
partially scaled state change $l'$ (corresponding to a state change $l$ in
the original chain) which occurred in the time interval $(0,u]$ and whose
intensity is given by 
\begin{equation}
\nu_{l}(\Upsilon \BN(u^-))= p\BN_{\Upsilon \BN(u^-),\Upsilon \BN(u^-)+l'} \label{intensity}
\end{equation}

In order to include the behavior on the boundary of the state space for the
components in $F$, we proceed as explained in the previous section.  The
set $C^{F}$ is split dynamically (depending on the current state of the
process $x$) into two subsets $\stella{C}^{F}_{x}$ and $\anello{C}^{F}_{x}$
defined as
\[
\stella{C}^{F}_{x}= \left\{ l \in C^{F} : \exists \; i \in B(x) \cap F \text{ such that either } l_{i} \neq 0 \text{ or } \frac{\partial p \BN_{x,x+l'}}{\partial x_{i}} \neq 0 \right\},
\]
and $\anello{C}^{F}_{x}=C^{F}-\stella{C}^{F}_{x}$, respectively.  Finally,
the hybrid switching jump diffusion process $\tilde{\Upsilon}\BN(u)$
accounting both for the hybrid definition of the model and for the boundary
conditions is given as
\begin{align}
d\tilde{\Upsilon} \BN_{j}(u) &= \sum_{l \in \anello{C}^{F}_{\tilde{\Upsilon}\BN(u)}} l_{j} f\left( \tilde{\Upsilon}\BN (u),l \right) du
+  \sum_{l \in \anello{C}^{F}_{\tilde{\Upsilon}\BN(u)}} \hspace{-1mm}  \frac{l_{j}}{\sqrt{N}} \sqrt{f \left( \tilde \Upsilon \BN(u),l \right) }dW_{l}(u)
+ \hspace{-2mm}\sum_{l \in C^{D} \cup \stella{C}^{F}_{\tilde{\Upsilon}\BN(u)}} \hspace{-1mm}\frac{l_{j}}{N}\;  dJ^{\BN}_{l}(u), \quad \quad j \in F, \nonumber\\
d\tilde \Upsilon \BN_{i}(u) &= \sum_{l \in C^{D}} \hspace{-1mm} l_{i}\;  dJ^{\BN}_{l}(u), \quad \quad i \in D,\label{hsjd}
\end{align}
where, again, $J\BN_{l}(u)$ is the process counting the events that cause
a partially scaled state change $l'$ (corresponding to a state change $l$
in the original chain) which occurred in the time interval $(0,u]$ and
whose intensity is given by \begin{equation}
\nu_{l}(\Upsilon \BN(u^-))= p\BN_{\Upsilon \BN(u^-),\Upsilon \BN(u^-)+l'}. \label{intensity2}
\end{equation}

The resulting process is a \emph{hybrid switching jump diffusion} in which
the jumps account both for the jumps of intrinsically discrete
components and for the jumps out of the boundaries of the fluid
components. Let us remark that \emph{hybrid} jump diffusions are not a
generalization of jump diffusion processes. On the contrary they can be
seen as a special case, since the discrete component can simply be
described as a jump-diffusion process that does only have jumps and no
diffusion.

Let us note that in the limit when $N\rightarrow \infty$ the noise in the
first equation of \eqref{hsjd} vanishes and the model becomes a PDMP in
which the discrete components modulates the ODE and the fluid levels have
an influence on the intensity \eqref{intensity2} of the jumps.  The
mathematical theory of hybrid switching diffusions together with the
special case of PDMP can be found in the recent monograph \cite{yin} where
also some hybrid switching jump-diffusions are briefly sketched in the
appendix.

\subsection{Simulation of the proposed hybrid switching jump diffusion} \label{algoritmoAparole}

From (\ref{hsjd}) an Euler scheme based, approximate simulation algorithm
can be derived directly.  Assume that the state of the process is known at
time $u$ and that we aim to simulate the process in a short time interval.
The length of this interval is $\Delta t$ if no discrete state change
occurs in $[u,u+\Delta t]$ and it is shorter otherwise.  First, we check if
a discrete state change occurs in the time interval $[u,u+\Delta t]$.  For
this purpose, we generate an exponential random variable whose rate,
$\nu$, is equal to the sum of the intensities of the involved counting
processes, i.e., $\nu=\sum_{l \in
C^{D} \cup \stella{C}^{F}_{\tilde{\Upsilon}\BN(u)}} \nu_{l}(\Upsilon \BN(u))$.
Let us denote this random variable by $r$.  If $r>\Delta t$ then no
discrete state change occurs in the considered interval.  If $r\leq
\Delta t$ then we have to simulate which counting process generates the
event.  This is done by generating a discrete random variable according to
a discrete probability distribution with mass function
$m_l=\nu_{l}(\Upsilon \BN(u))/\nu$ with $l \in
C^{D} \cup \stella{C}^{F}_{\tilde{\Upsilon}\BN(u)}$. We denote the
resulting random variable by $k$.  In order to simulate the diffusion we
need to generate $|\anello{C}^{F}_{\tilde{\Upsilon}\BN(u)}|$ random
variables according to the standard normal distribution.  Let us denote
these random variables by $W'_l$ with
$l \in \anello{C}^{F}_{\tilde{\Upsilon}\BN(u)}$.  Having generated the
above random numbers the state of the process is updated first according to
\begin{align}
\nonumber
\tilde{\Upsilon} \BN_{j}(u+\min(r,\Delta t)) = & \tilde{\Upsilon}  \BN_{j}(u)+
\sum_{l \in \anello{C}^{F}_{\tilde{\Upsilon}\BN(u)}} l_{j} f\left( \tilde{\Upsilon}\BN (u),l \right) \min(r,\Delta t)
+  \sum_{l \in \anello{C}^{F}_{\tilde{\Upsilon}\BN(u)}}   \frac{l_{j}}{\sqrt{N}} \sqrt{f \left( \tilde \Upsilon \BN(u),l \right)}
W'_{l} \sqrt{\min(r,\Delta t)}
+ \\ & I(r\leq \Delta t) \frac{l_{k}}{N} \quad \quad ~~~~~~~~~~~~~~ j \in F \nonumber\\
\tilde \Upsilon \BN_{i}(u+ \min(r,\Delta t))
= &  \Upsilon \BN_{i}(u) + I(r\leq \Delta t) l_{k} \quad \quad i \in D \nonumber
\end{align}
where $I$ is the indicator function.  Due to the presence of the diffusion,
it is possible that the previous update results in a state in which some
components are smaller than their possible minimal or larger than their
possible maximal values.  If a component is smaller (greater) than its
minimal (maximal) value then we set it to its minimal (maximal) value.  In
other words, we simulate the process in $[u,u+\Delta t]$ as if it was
unbounded and subsequently we test whether the process left its valid state
space.  If $\Delta t$ is sufficiently small then the above simulation
procedure results in traces that reflect properly the behavior of the
original process.  The choice of $\Delta t$ can be made on the basis of the
actual state of the process by taking into account the maximum among the
drifts.  We found that for our examples $\Delta t=\max\left\{ l_{j}
f\left( \tilde{\Upsilon}\BN (u),l \right)~\mbox{with}~
l \in \anello{C}^{F}_{\tilde{\Upsilon}\BN(u)} \right\}^{-1}/100$ is a
reasonable compromise between execution time and precision.  A more
implementation oriented description of the simulation algorithm is provided
in \ref{app:algo}.

Note that with this simulation approach the borders are reached with delay.
For instance, if a component is positive at time $u$ and it is updated to a
negative value at time $u+\Delta t$, then we assume that the process
reached level 0 at time $u+\Delta t$.  It can also happen that we miss the
fact that the process reached a border.  Indeed there is a positive
probability that an unbounded diffusion process is positive at time $u$ and
at time $u+\Delta t$ and it is negative somewhere inside $[u,u+\Delta]$.
With sufficiently small $\Delta t$ this causes insignificant imprecision.
A more precise treatment of the borders, with which the time when the
process reaches the border is estimated more precisely, is possible and we
plan to experiment with it in the future.

Let us remark that the simulation algorithm described above is similar to
the hybrid stochastic simulation methods based on dynamic partitioning
(e.g., \cite{griffith, salis2}).  The main difference is that, while these
hybrid simulation methods were derived heuristically through clever
simulation speed ups, our simulation algorithm arises by applying a simple
Euler discretization scheme to the SDEs \eqref{hsjd} that describe the
approximating HSJD model.

%
%

\section{Analytic calculation\label{Sec:analytic}}

In the previous section we proposed a stochastic process, namely a hybrid
switching jump diffusion (HSJD), to approximate a class of CTMCs.  In the
case when the system contains a single fluid component,
the
process \eqref{hsjd} reduces to the \emph{elementary return process}
studied by Feller in\footnote{The elementary return process is described in
details in \cite{feller1d}, but be aware of the numerous misprints in the
formulae of that paper. A less readable, but misprint free version is
contained in \cite{fellerParabolic}.} \cite{feller1d, fellerParabolic}.  In
this section we illustrate the possibility of calculating the distribution
of the involved quantities by numerical integration of the Fokker-Plank
equations there introduced.  To this end, we first analyze a simple model
without switching and then extend it with a switch.

Let us consider the so called \emph{crazy clock reaction} (cf. \cite{ErdiLente} and references quoted therein). It is an autocatalytic system composed of two reactions, $A \rightarrow B$ and $A+B
\rightarrow 2B$, and  initial state  $(N,0)$ corresponding to $N$ molecules of type A and $0$ molecules of type B.
The intensities
associated with the two transitions in state $(i,j)$ are $\lambda_1 i$ and
$\lambda_2 i j/N$, respectively.
The second reaction is usually much faster since it has a quadratic rate, but it cannot occur until the first molecule of $B$ has been produced by means of the first reaction.

As both reactions transform one $A$ into one $B$, we have the invariant
$i+j=N$ and hence a single variable is sufficient to describe a state.  In the
following we will use the number of $A$s as state descriptor.  In the
associated CTMC the intensity of the possible transitions is
$q_{(i),(i-1)}=\lambda_1 i+\lambda_2 i (N-i)/N$.  Accordingly, the function
$f$ required by Definition~\ref{def:density_dependent} is
\[
f(x,l)=\left\{
\begin{array}{ll}
 \lambda_1 x + \lambda_2x(1-x)&\mbox{if}~l=-1 \\
 0 & \mbox{otherwise}
\end{array}
\right.
\]
We evaluate the model with $\lambda_1=3, \lambda_2=6000$ and $N=1000$. Note
that after normalization the variable $x=\frac{i}{N}$ is between $0$ and $1$.

Since there is no switch in the model, the HSJD approximation can be derived
from (\ref{eq:Jsde}), assuming the following form
\begin{equation}
\begin{split}
\label{eq:m1}
d \tilde{Y}^{\BN}(u) =
- I\left(0<\tilde{Y}^{\BN}(u)<1\right)\left( f \left( \tilde{Y}^{\BN}(u),-1 \right) du
+ \frac{1}{\sqrt{N}} \sqrt{f \left( \tilde{Y} ^{\BN}(u),-1 \right) }dW_{-1}(u)\right)\\
- I\left(\tilde{Y}^{\BN}(u)=1 \text { or } \tilde{Y}^{\BN}(u)=0\right) \frac{1}{N} dM^{\BN}(u)
\end{split}
\end{equation}
where $I$ is the indicator function and the intensity of the counting
process, $M^{\BN}(u)$, is $Nf(\tilde{Y}^{\BN}(u),-1)$, that is $N\lambda_1$
when $\tilde{Y}^{\BN}(u)=1$ and $0$ when $\tilde{Y}^{\BN}(u)=0$.  In
(\ref{eq:m1}) the first term on the right hand side accounts for the drift and for the diffusion which is ``switched off'' when the process is at the
upper boundary.  The second term is the jump process that moves the
process away from the upper boundary.  The initial condition is
$\tilde{Y}^{\BN}(0)=1$.  Let us remark again that at the lower boundary
$f(0,-1)=0$ and hence the intensity of the jumps is zero so that the lower
boundary is absorbing.

The probability density function (pdf) of the quantity of $A$s at time $u$
is mixed, it has a probability mass at the boundaries (at $x=0$
and at $x=1$) and it is continuous elsewhere.  The continuous part will be
denoted by $\pi(u,x)$, i.e.,
\[
\pi(u, x)=\frac{\partial}{\partial x}\mathbb{P}\left\{\tilde{Y}^{\BN}(u)\leq
x| \tilde{Y}^{\BN}(0)= 1\right\} ~\mbox{for}~ 0<x<1
\]
We will refer to the probability masses as $\pi_{0}(u)$ and $\pi_{1}(u)$,
i.e.,
\[
\pi_{0}(u)=\mathbb{P}\left\{\tilde{Y}^{\BN}(u)= 0| \tilde{Y}^{\BN}(0)= 1\right\}, ~~
\pi_{1}(u)=\mathbb{P}\left\{\tilde{Y}^{\BN}(u)= 1| \tilde{Y}^{\BN}(0)= 1\right\}, ~~
\]
Furthermore, we will denote by $\pi(u,0^+)$ and $\pi(u,1^-)$ the limit values of
the continuous density at 0 and 1, respectively.

\begin{figure}
\begin{minipage}{0.48\textwidth}
\includegraphics[width=0.9\textwidth]{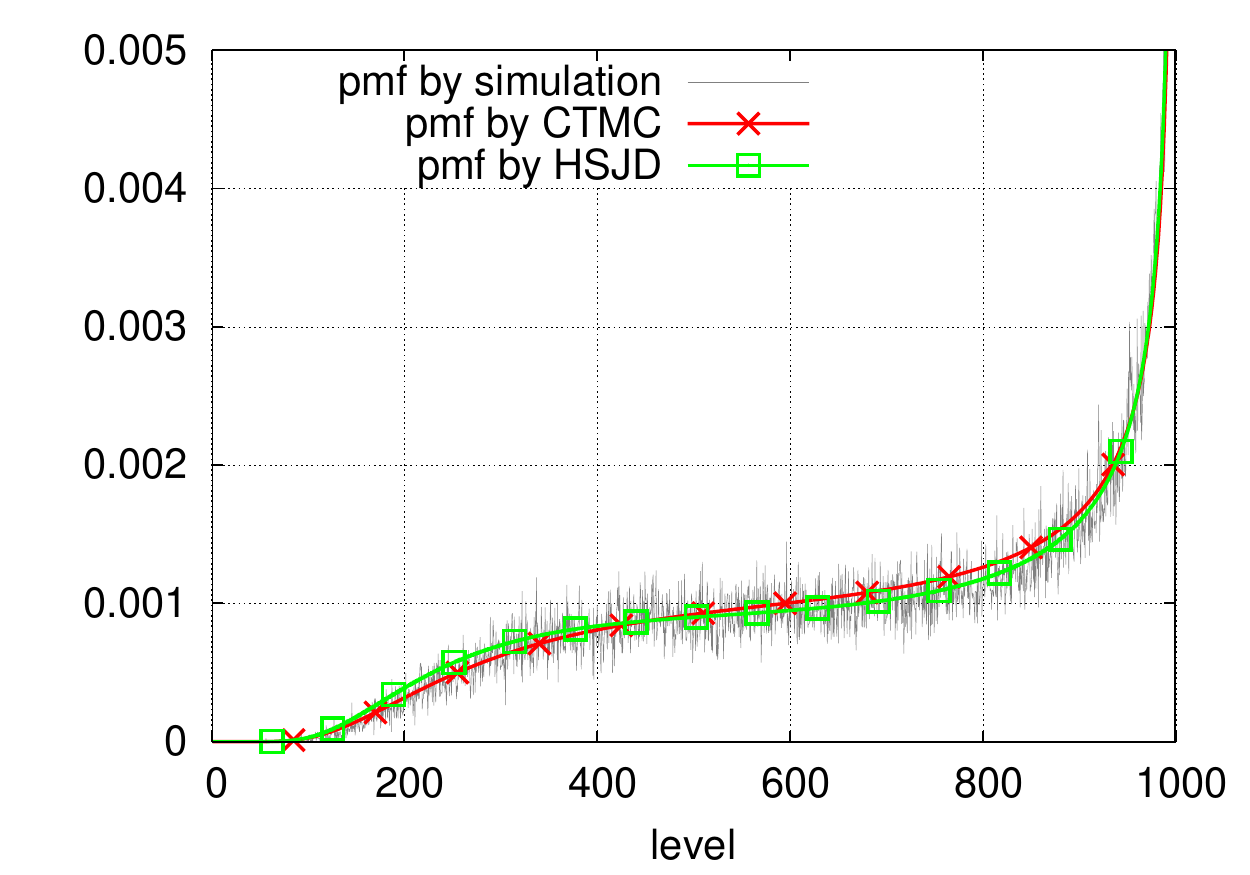}
\end{minipage}
\hfill
\begin{minipage}{0.48\textwidth}
\includegraphics[width=0.9\textwidth]{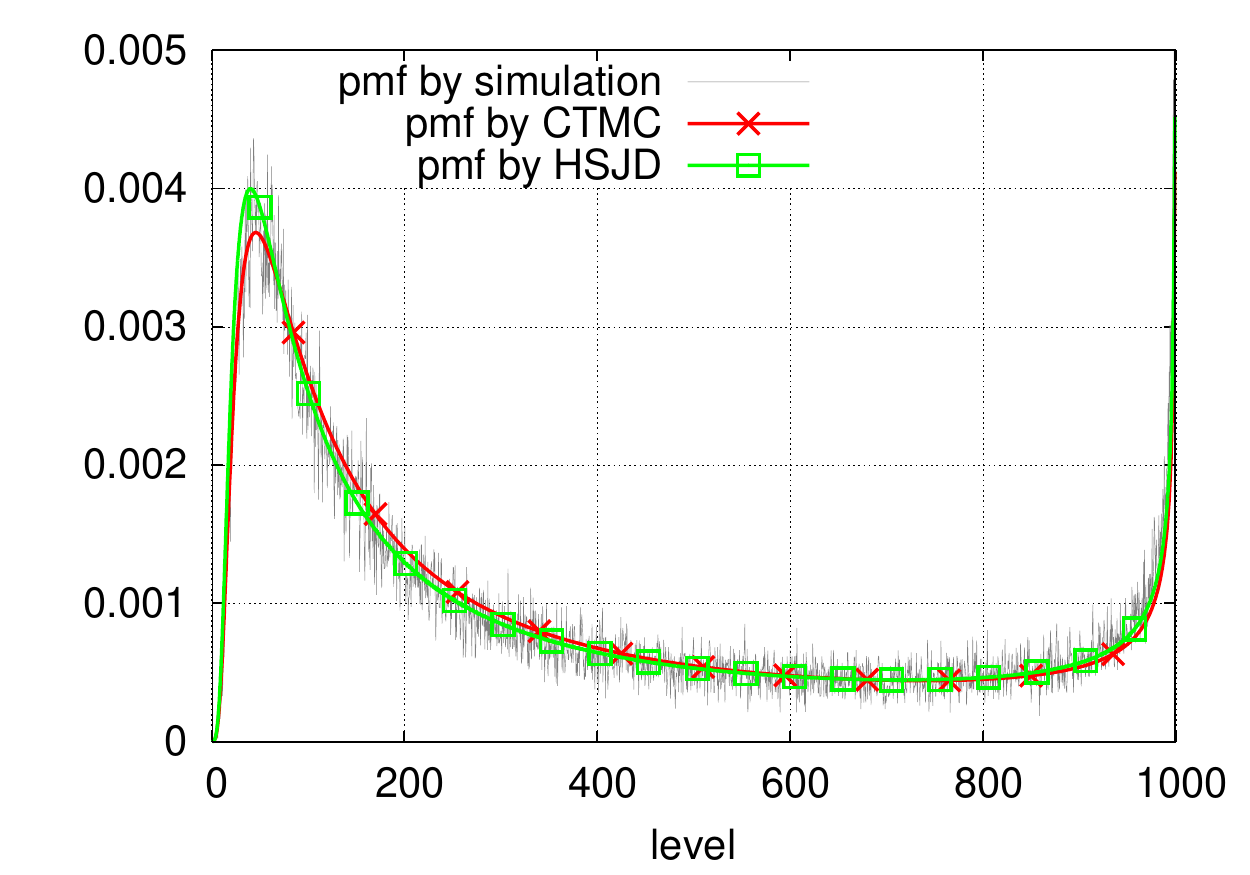}
\end{minipage}
\caption{Probability mass function of the quantity of $A$s in the model $A
  \rightarrow B, A+B \rightarrow 2B$, at time $u=0.0012$ (left) and at
  time $u=0.0016$ (right)\label{fig:a1}}
\end{figure}

\begin{figure}
\begin{center}
\begin{minipage}{0.48\textwidth}
\includegraphics[width=0.9\textwidth]{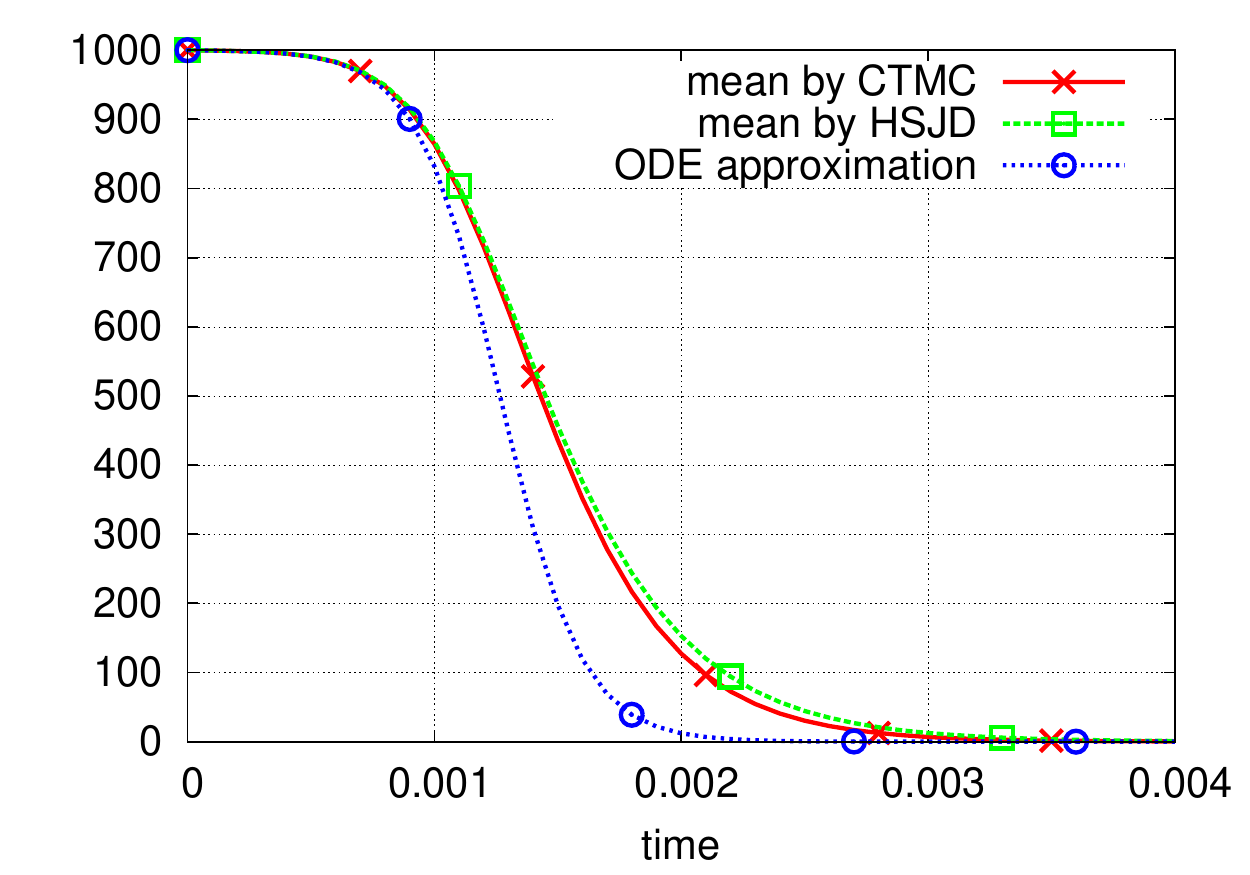}
\end{minipage}
\end{center}
\caption{Average quantity of $A$s in the model $A
  \rightarrow B, A+B \rightarrow 2B$, as function of time
\label{fig:mean}}
\end{figure}

The time evolution of $\pi(u,x)$ is described by a
Fokker-Planck partial differential equation (PDE) equipped with the suitable
boundary conditions \cite{feller1d}.  We have
\begin{equation}
\label{eq:fp}
\frac{\partial}{\partial u}\pi(u,x)=
\frac{\partial}{\partial x} \left(f \left(x,-1\right) \pi(u,x) \right)+
\frac{\partial^2}{\partial x^2} \left(\frac{f(x,-1)}{2N} \pi(u,x) \right)+
\delta\left(1-\frac{1}{N}\right)Nf(1,-1)\pi_{1}(u) ~~~~~~ \mbox{for}~ 0<x<N
\end{equation}
where $\delta(y)$ denotes a Dirac delta distribution centered at $y$.  In
(\ref{eq:fp}) on the right hand side, the first term corresponds to the
drift of the process, the second to the diffusion coefficient (and these
two are standard in Fokker-Planck equations) while the third describes the
way the process ``jumps away'' from the upper boundary.  This last term
contains a Dirac distribution because the level reached after the jump has a
deterministic distribution.  It takes into account also the intensity of
the corresponding rate of the original CTMC, $Nf(N,-1)$, and depends
on the probability mass at $x=1$ as well.  The boundary condition at the upper
boundary is given by
\[
\frac{\partial}{\partial u}\pi_{1}(u)=
-f \left(1,-1\right) \pi(u,1^-)-
\frac{\partial}{\partial x} \left.\left(\frac{f(x,-1)}{2N} \pi(u,x) \right)\right|_{x=1^-}-
Nf(1,-1)\pi_{1}(u)
\]
where the first two terms provide the rate of net probability flux toward
the upper boundary at time $t$ and the third term is the flux exiting from
the boundary due to the jumps.  At the lower boundary we have $f(0,-1)=0$
which means that both the drift and the diffusion coefficient are 0.
Nevertheless, thanks to the diffusion the lower boundary is reachable.
This boundary is absorbing and the probability mass can be calculated
simply by
\[
\pi_{0}(u)=1-\pi_{1}(u) -\int_{0}^{1} \pi(u,x)dx
\]
The initial condition is
\[
\pi_{0}(u)=0 \qquad \pi_{1}(u)=1  \qquad  \pi(0,x)=0 \text{,  for every }0<x<N
\]
which means that the initial level is 1 with probability 1 (note that we
deal with the normalized process).  Since both boundaries are regular
(cf. \cite{feller1d}), to single out the correct solution we need to impose
the following further boundary conditions
\[
 \pi(u,1^-)=0\qquad \text{   and   }\qquad   \lim_{x\rightarrow 0^{+}} x^{2} \pi(u,x)=0.
\]
The PDE in (\ref{eq:fp}) together with the boundary and initial conditions
can be solved numerically by discretizing the involved variables ($u$ and
$x$) and by applying a finite volume scheme.  In Figure~\ref{fig:a1} we
depict the distribution of the unnormalized quantity of $A$s for two time points.  The
figures show three probability mass functions (pmf).  The first one was
obtained by solving the original CTMC, the second was calculated by the
finite volume scheme applied to the PDE given in (\ref{eq:fp}) and the
third was obtained by simulating the HSJD given in (\ref{eq:m1}).  We
created $10^6$ simulation traces, which took about two hours to complete, and still insufficient to produce a smooth pmf.  The finite volume scheme for the PDE was
implemented in scientific python and the calculations took about 10
seconds.  There is a good agreement between the results based on simulation
and those based on the PDE.  The comparison against the pmf obtained based
on the CTMC reveals that the HSJD is a good approximation of the original
process.  In Figure~\ref{fig:mean} we depicted the average amount of $A$s
obtained from the original CTMC, from the HSJD approximation (using the
PDE) and from the ODE approximation.  There is a large time interval where
the ODE approximation gives largely imprecise idea of the amount of $A$s in
the system.  At time $u=0.002$ the ODE curve is at about 12 while the mean in
the original model is at about 127.  

The explanation for this behavior of the ODE approximation is that in the
CTMC and in the HSJD the second reaction ($A+B
\rightarrow 2B$) is inhibited up to the point when the
other reaction ($A \rightarrow B$) takes place for the first time. On the
contrary, in the ODE model the terms of the equations corresponding to the
two reactions both start to contribute at the very beginning of the
evolution of the model, resulting in an anticipated decay. This
interpretation is confirmed by Fig.~\ref{fig:a1}: at time $u=0.0016$ the
increase of the pmf toward 1000 indicates that there are realizations in
which no reaction has taken place yet or the first took place shortly
before.

We extend now the model with a one-way switch that modulates the speed of the
reaction $A+B \rightarrow 2B$.  The switch is represented by a third
compound denoted by $C$.  We have now three reactions:
\[
A \rightarrow B, ~~~ A+B \rightarrow 2B, ~~~ C \rightarrow \emptyset
\]
and the corresponding initial state is $(N,0,1)$.  In state $(i,j,k)$ the intensity of
the first reaction is $\lambda_1 i$, the intensity of the second reaction
is $\lambda_2 i j/N$ if $k=1$ and it is $2 \lambda_2 i j/N$ if $k=0$ (i.e.,
the second reaction is two times faster if there is no $C$ in the system),
and the intensity of the third reaction is $c(i)=\max(0, \lambda_3 (i -
(N - S))/S)$.  This means that the third reaction can occur only if
$i>(N-S)$, its intensity is a linear function of $i$ that reaches its maximum value $\lambda_3$ at $i=N$.  As before we have $i+j=N$ and
hence two variables, $i$ and $k$, are sufficient to describe a state.  In the
associated CTMC the intensities of the possible transitions are
$q_{(i,1),(i-1,1)}=\lambda_1 i+\lambda_2 i (N-i)/N$,
$q_{(i,0),(i-1,0)}=\lambda_1 i+2 \lambda_2 i (N-i)/N$ and
$q_{(i,1),(i,0)}=c(i)$.  We will apply a HSJD approximation in which the
number of $A$s is described by a continuous quantity and the number of $C$s
is maintained discrete.  This leads to a jump diffusion with a switch.  We
will refer as mode 1 (mode 2) the situation in which the number of $C$s
in the system is 1 (0).  In both mode 1 and 2, the rate of the reaction that changes
the number of $A$s is in density dependent form.  The corresponding $f$
function of Definition~\ref{def:density_dependent} is
\[
f(x,l)=\left\{
\begin{array}{ll}
 \lambda_1 x_1 + \lambda_2x_1(1-x_1)&\mbox{if}~l=(-1,0)~\mbox{and}~x_2=1 \\
 \lambda_1 x_1 + 2\lambda_2x_1(1-x_1)&\mbox{if}~l=(-1,0)~\mbox{and}~x_2=0 \\
 0 & \mbox{otherwise}
\end{array}
\right.
\]
Note that after normalization of the fluid states the quantity
$x_1=\frac{i}{N}$ ranges from 0 to 1.  The HSJD approximation is given now in
(\ref{hsjd}).  For the continuous part we have
\begin{equation}
\begin{split}
\label{eq:m2}
d \tilde{Y_1}^{\BN}(u) =
- I\left(0<\tilde{Y_1}^{\BN}(u)<1\right)\left( f \left( \tilde{Y}^{\BN}(u),(-1,0) \right) du
+ \frac{1}{\sqrt{N}} \sqrt{f \left( \tilde{Y} ^{\BN}(u),(-1,0) \right) }dW_{-1}(u)\right)\\
- I\left(\tilde{Y_1}^{\BN}(u)=1 \text { or } \tilde{Y_1}^{\BN}(u)=0\right) \frac{1}{N} dM^{\BN}(u)
\end{split}
\end{equation}
which is almost identical to (\ref{eq:m2}) because in this model the
transitions that change the discrete component do not change the
continuous component.  For the discrete component
\begin{equation}
\label{eq:m2b}
d \tilde{Y_2}^{\BN}(u) = - I\left(\tilde{Y_2}^{\BN}(u)=1\right)  dJ^{\BN}(u)
\end{equation}
where $J^{\BN}(u)$ is counting the occurrences of the reaction
$C \rightarrow \emptyset$ with intensity $c\left(\tilde{Y_1}^{\BN}(u)\right)$. The
initial condition is $\tilde{Y}^{\BN}(0)=(1,1)$.

In order to describe the transient behavior of the model we need to refer
to the pdf of the quantity of $A$s in mode 1 and in mode 2.  The continuous
part of these densities will be denoted as
\[
\pi_1(u, x)=\frac{\partial}{\partial x}\mathbb{P}\left\{\tilde{Y_1}^{\BN}(u)\leq
x,\tilde{Y_2}^{\BN}(u)=1 | \tilde{Y}^{\BN}(0)= (1,1)\right\}, ~~~
\pi_2(u, x)=\frac{\partial}{\partial x}\mathbb{P}\left\{\tilde{Y_1}^{\BN}(u)\leq
x,\tilde{Y_2}^{\BN}(u)=0 | \tilde{Y}^{\BN}(0)= (1,1)\right\}
\]
with $0<x<1$, while the overall pdf of the quantity of $A$s is \[\pi(u, x)=\pi_1(u, x)+\pi_2(u, x).\]  The mode-specific masses at the boundaries will be referred to as
\[
\pi_{1,0}(u)=\mathbb{P}\left\{\tilde{Y}^{\BN}(u)= (0,1)| \tilde{Y}^{\BN}(0)= (1,1)\right\}, ~~
\pi_{1,1}(u)=\mathbb{P}\left\{\tilde{Y}^{\BN}(u)= (1,1)| \tilde{Y}^{\BN}(0)= (1,1)\right\}
\]
and
\[
\pi_{2,0}(u)=\mathbb{P}\left\{\tilde{Y}^{\BN}(u)= (0,0)| \tilde{Y}^{\BN}(0)= (1,1)\right\}, ~~
\pi_{2,1}(u)=\mathbb{P}\left\{\tilde{Y}^{\BN}(u)= (1,0)| \tilde{Y}^{\BN}(0)= (1,1)\right\}
\]
while their overall counterparts are
\[\pi_{0}(u)=\pi_{1,0}(u)+\pi_{2,0}(u)\qquad \text{ and }\qquad \pi_{1}(u)=\pi_{1,1}(u)+\pi_{2,1}(u).\]
The Fokker-Planck equation that describes the evolution of the density in
mode 1 is
\begin{align}
\label{eq:fp2}
\frac{\partial}{\partial u}\pi_1(u,x)= &
\frac{\partial}{\partial x} \left(f \left((x,1),(-1,0)\right) \pi_1(u,x) \right)+
\frac{\partial^2}{\partial
x^2} \left(\frac{f((x,1),(-1,0))}{2N} \pi_1(u,x) \right)+\\ & \notag
\delta\left(1-\frac{1}{N}\right)Nf((1,1),(-1,0))\pi_{1,1}(u)
-c(x)\pi_1(u,x) ~~~~~ 0<x<1
\end{align}
where the last term is due to the switch.  Similarly, in mode 2 we have
\begin{align}
\label{eq:fp3}
\frac{\partial}{\partial t}\pi_2(u,x)= &
\frac{\partial}{\partial x} \left(f \left((x,0),(-1,0)\right) \pi_2(u,x) \right)+
\frac{\partial^2}{\partial
x^2} \left(\frac{f((x,0),(-1,0))}{2N} \pi_2(u,x) \right)+\\ & \notag
\delta\left(1-\frac{1}{N}\right)Nf((1,0),(-1,0))\pi_{2,1}(u)
+c(x)\pi_1(u,x) ~~~~~ 0<x<1
\end{align}
Also the boundary conditions at 1 contain a term that takes into account the
switch of the model.  They are
\begin{align}
\notag
\frac{\partial}{\partial t}\pi_{1,1}(u)=&
-f \left((1,1),(-1,0)\right) \pi_1(u,1^-) -
\frac{\partial}{\partial
x} \left.\left(\frac{f((x,1),(-1,0))}{2N} \pi_1(u,x) \right)\right|_{x=1^-}
\\ \notag & -
Nf((1,1),(-1,0))\pi_{1,1}(u)-c(1)\pi_{1,1}(u)
\end{align}
and
\begin{align}
\notag
\frac{\partial}{\partial t}\pi_{2,1}(u)=&
-f \left((1,0),(-1,0)\right) \pi_2(u,1^-) -
\frac{\partial}{\partial
x} \left.\left(\frac{f((x,0),(-1,0))}{2N} \pi_2(u,x) \right)\right|_{x=1^-}
\\ \notag & -
Nf((1,1),(-1,0))\pi_{2,1}(u)-c(1)\pi_{1,1}(u)
\end{align}
For the boundary at 0, conditions similar to those reported
for the model without switch  can be written (note that the intensity of the switch is 0 at
the lower boundary).  This formulation does not allow one to distinguish the
mass at 0 in mode 1 and the mass at 0 in mode 2.  It is possible to
distinguish these two masses, but this would require a slightly different
treatment of the lower boundary.

\begin{figure}
\begin{minipage}{0.48\textwidth}
\includegraphics[width=0.9\textwidth]{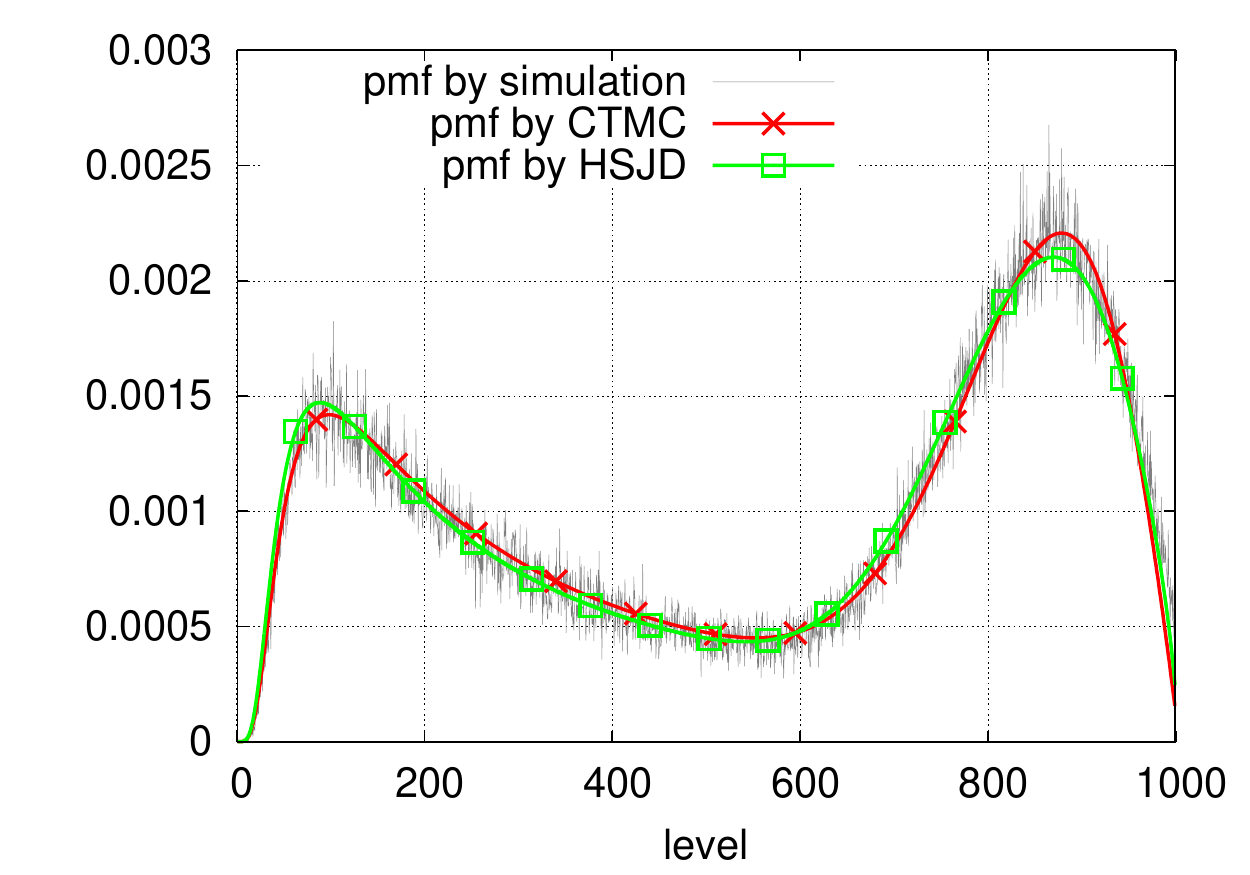}
\end{minipage}
\hfill
\begin{minipage}{0.48\textwidth}
\includegraphics[width=0.9\textwidth]{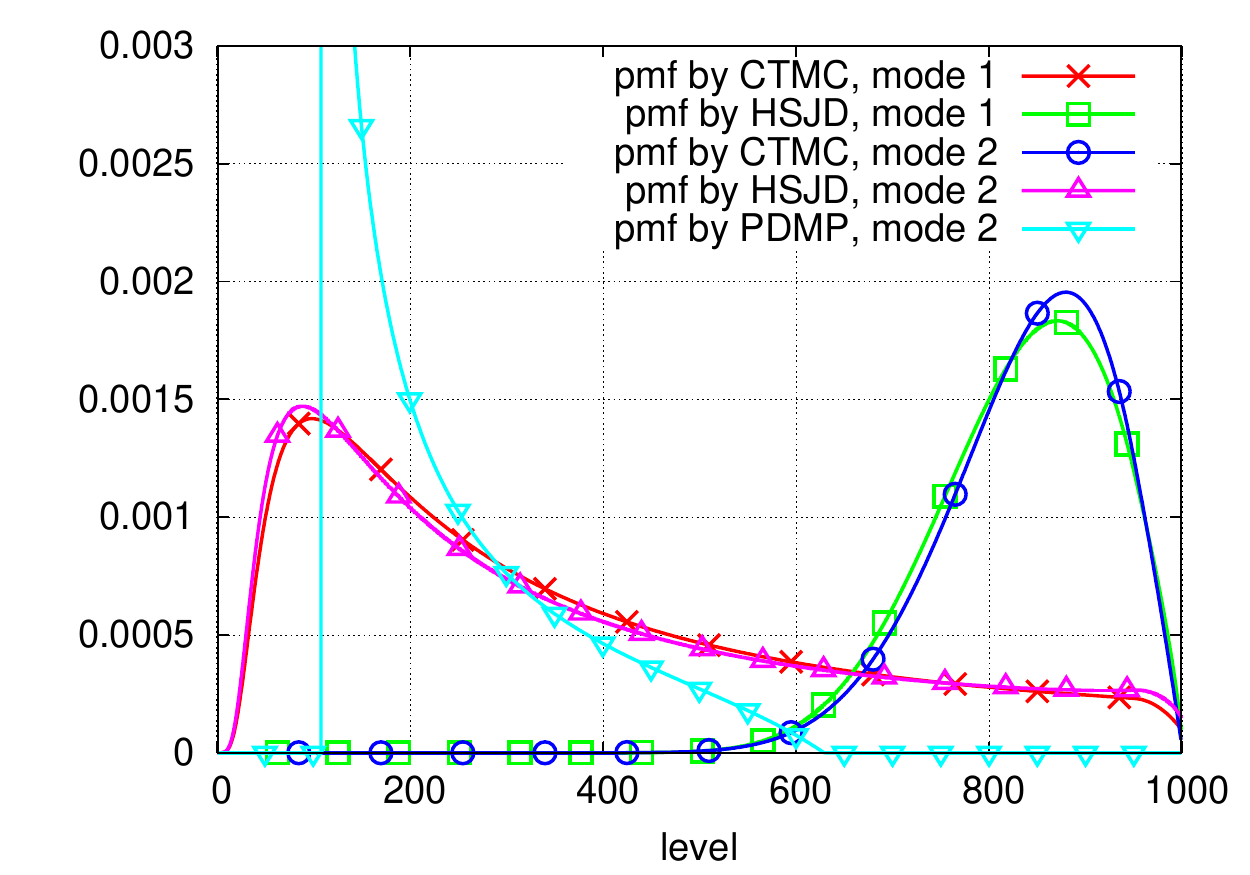}
\end{minipage}
\caption{Distribution of the quantity of $A$s at time $u=0.003$; overall pmf
 (left) and pmf per mode  (right). \label{fig:b1}}
\end{figure}

We considered the model with the switch with the following parameters (in the unnormalized scale):
$\lambda_1=3, \lambda_2=3000, \lambda_3=500, S=50$ and $N=1000$.  We
analyzed the model by the method of finite volumes and the execution times
are similar to those reported for the model without switch.  The
distribution of the unnormalized quantity of $A$s at $u=0.003$ is depicted in
Figure~\ref{fig:b1}.  Also in this case, there is a good agreement among
the numerical solution of the PDEs, the simulation based results and the
behavior of the original CTMC. In this example an approach based on pure ODEs is not feasible due to the presence of the switch, while a hybrid approach with a PDMP that switches between two ODEs is reasonable. On the right side of Figure~\ref{fig:b1} we
depicted the distribution obtained by the PDMP approach as well.
In mode 1 the PDMP approach leads to a single probability mass at
about 848 because this approach provides a deterministic description inside
each mode.  This mass in not depicted in the figure.  In mode 2 the PDMP approach gives a distribution because the time point at which the
switch occurs is random.  The support of this distribution is
determined by the minimal and maximal times at which the switch changes
from mode 1 to mode 2.  The minimal time is 0 and in this case the level in
mode 2 at $u=0.003$ is about 109.  The maximal time is about 0.0022 because
in mode 1 the level is $N-S=950$ at this time point.  If
the switch snaps at $u=0.0022$ then the level in mode 2 at $u=0.003$ is about
629.  On the contrary, with the HSJD approach the level in mode 2 can be
any value in the interval [0,1000]. The HSJD approach provides a much better approximation of the CTMC with respect to the PDMP one.

\section{Simulation based results}\label{Sec:res}

%
%
%

In this section we analyze two biological models to provide a comparison of
the quality and robustness of the approximation that we have proposed, with
respect to more standard approaches. The pure jump-diffusion approximation
(all quantities are fluid except at the boundaries) is referred to as SDE
approximation both in the figures and in the text, while the proposed
hybrid switching jump diffusion as HSDE (Hybrid SDE).  The deterministic
approximation is referred to as ODE approximation and our comparisons
involve also a PDMP approach which is referred to as HODE (Hybrid ODE).

The first model, which represents a viral intracellular kinetics, shows that
\begin{itemize}
\item the SDE approximation is more informative and accurate than
  that based on ODEs when there is a non-negligible probability to find the
  original discrete process on the barriers and the quantities under study
  have  bi-modal behaviors;
\item although the approximation provided by the SDE is less accurate than
  that generated by using a HODE approach, the jump-diffusion is more accurate
  on the barriers;
\item the HSDE approach is the most accurate among all the methods
  discussed in this paper.
\end{itemize}
The second and more complex model, based on a transcription regulation
phenomenon, is introduced to illustrate more clearly the last item of the
previous list.  This cannot be done with the first model because it is
characterized by an exceedingly simple dynamic behavior.

The comparison of the different approaches has been carried out with a
prototype implementation of the algorithm described in Section
\ref{algoritmoAparole} and \ref{app:algo}, integrated in the GreatSPN
framework~\cite{BabarBDM10}. A more detailed description of the tool is
given in Section \ref{tool}. The results computed by our prototype
implementation have been processed through the R framework \cite{Rurl} to
derive statistical information and graphics. All the results have been
obtained on a 2.13 GHz Intel I7 processor with 8GB of RAM.

\subsection{Viral intracellular kinetics}

The first model, describes the intracellular kinetics of a generic virus and has been studied
in~\cite{srivastava}.
It is described by the following six reactions:
\begin{eqnarray}
\notag
gen&\stackrel{k_1}{\rightarrow}& tem\\
\notag
tem	&\stackrel{k_2}{\rightarrow}& \emptyset \\
\notag
tem	&\stackrel{k_3}{\rightarrow}&tem+gen \\
\notag
gen+struct&\stackrel{k_4}{\rightarrow}& \emptyset \\
\notag
tem	&\stackrel{k_5}{\rightarrow}& tem+struct \\
struct &\stackrel{k_6}{\rightarrow}& \emptyset
\end{eqnarray}
where \emph{gen} represents the genomic viral nucleic acids, \emph{tem} the
template of viral nucleic acid transcribed to synthesize every viral
component, and \emph{struct} the viral structural protein.  In details,
reaction $k_1$  models the
integration of the genomic viral nucleic acids  into
the host genome to form templates.  Furthermore \emph{gen} can be packaged
(i.e. reaction $k_4$) within structural proteins  to form progeny virus as described by the fourth reaction.
After the initial virus infection, the amplification of the viral template
is modeled by reaction $k_3$. Then,
the synthesis of the viral structural protein is represented by reaction
$k_5$.  Finally, reactions $k_2$ and $k_6$ represent the degradation of
\emph{tem} and \emph{struct}, respectively.
The corresponding SPN  model is shown in Figure~\ref{Fig:Exp1} of ~\ref{App:1}.

As shown in~\cite{srivastava} using linear stability analysis, the system exhibits two equilibriums:
one in which all the components are null is unstable, while the other is stable.
Hence, initializing the system close to the unstable equilibrium we can observe that the
ODE-based methods always reaches the \emph{stable} equilibrium, while
stochastic simulation methods can reach also the \emph{unstable}
equilibrium with non-negligible probability.

We computed the transient behavior of the model along a time interval that
extends from $0$ to $200$ days using Monte Carlo simulation, ODE, SDE,
HODE, and HSDE under the assumptions that a single molecule of \emph{tem}
is present in the system at the beginning of the analysis and that the
reaction kinetic constants are those reported in Table~\ref{tab:exp1rates}.
The hybrid approaches consider \emph{gen} and \emph{tem} as discrete
quantities, see \cite{menz}.

\begin{table}
\centering
\begin{tabular}{|c c c c c c c c c c |}
\hline
$\mu_1$ & $\mu_2$ &$\mu_3$ &$\mu_4/N$ &$\mu_5$ &$\mu_6$ &\emph{gen} & \emph{tem} & \emph{struct} & N\\
\hline
$0.025$ & $0.25$ & $1.0$ &$7.5 \times 10^{-6}$ & $1000$ & $2.0$& 0 & 1 &0& 20000\\
\hline
\end{tabular}
\caption{Rates and initial condition  of the virus model in $days^{-1}$.}\label{tab:exp1rates}
\end{table}

\begin{figure}[tbp]
   \centering
   \includegraphics[width=.50\textwidth]{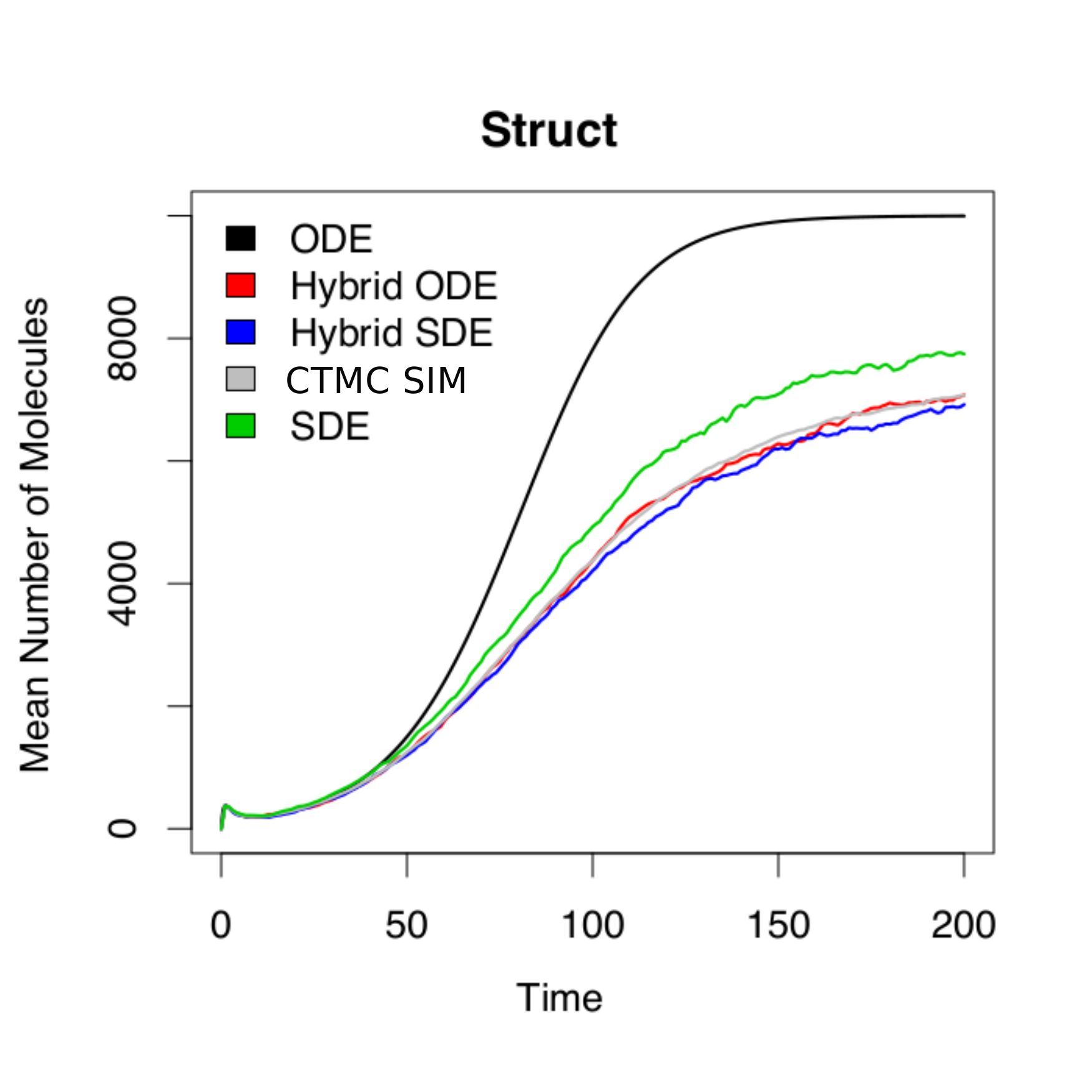}
   \caption{Mean of the number of \emph{struct} moleculas as function of the time. Comparison between Monte Carlo simulation, ODE, HODE, SDE, HSDE.}\label{Fig:1}
\end{figure}

Figure \ref{Fig:1} reports a first comparison between the simulation of the
original model and the approximations obtained with the four fluid or hybrid
approximations. The comparison is performed by depicting the expected
number of molecules of \emph{struct} as a function of time. Observe that
confidence intervals for Monte Carlo simulation, SDE, and HODE are not
explicitly reported on these diagrams to make them more readable;
nevertheless these results have been computed using 5000 runs to insure a
high level of confidence.  The ODE approach deviates quite soon from the
trajectory obtained from the simulation of the original process. In
particular, it overestimates the expected number of molecules of \emph{tem}
by flattening around 10000, while the simulation suggests that this limit
should be around 7000.  Such inaccuracy is mainly due to the initial low
number of reactants (i.e. \emph{tem}), typical of this type of systems.
Indeed, due to their deterministic nature, the ODEs cannot fall into the
unstable equilibrium state unless they are initialized exactly there, while the
CTMC can jump to the state for which the infection is blocked.
Although the difference with the original trajectory is still substantial,
SDE is much more accurate than ODE; the error between the curve
generated using the SDE approach and that obtained by simulating the CTMC
is confined within a $10\%$ level.  The HODE and HSDE approaches are
instead capable of reproducing the mean behavior of the original process in
a satisfactory manner.

\begin{figure}[!h]
\includegraphics[width=\textwidth]{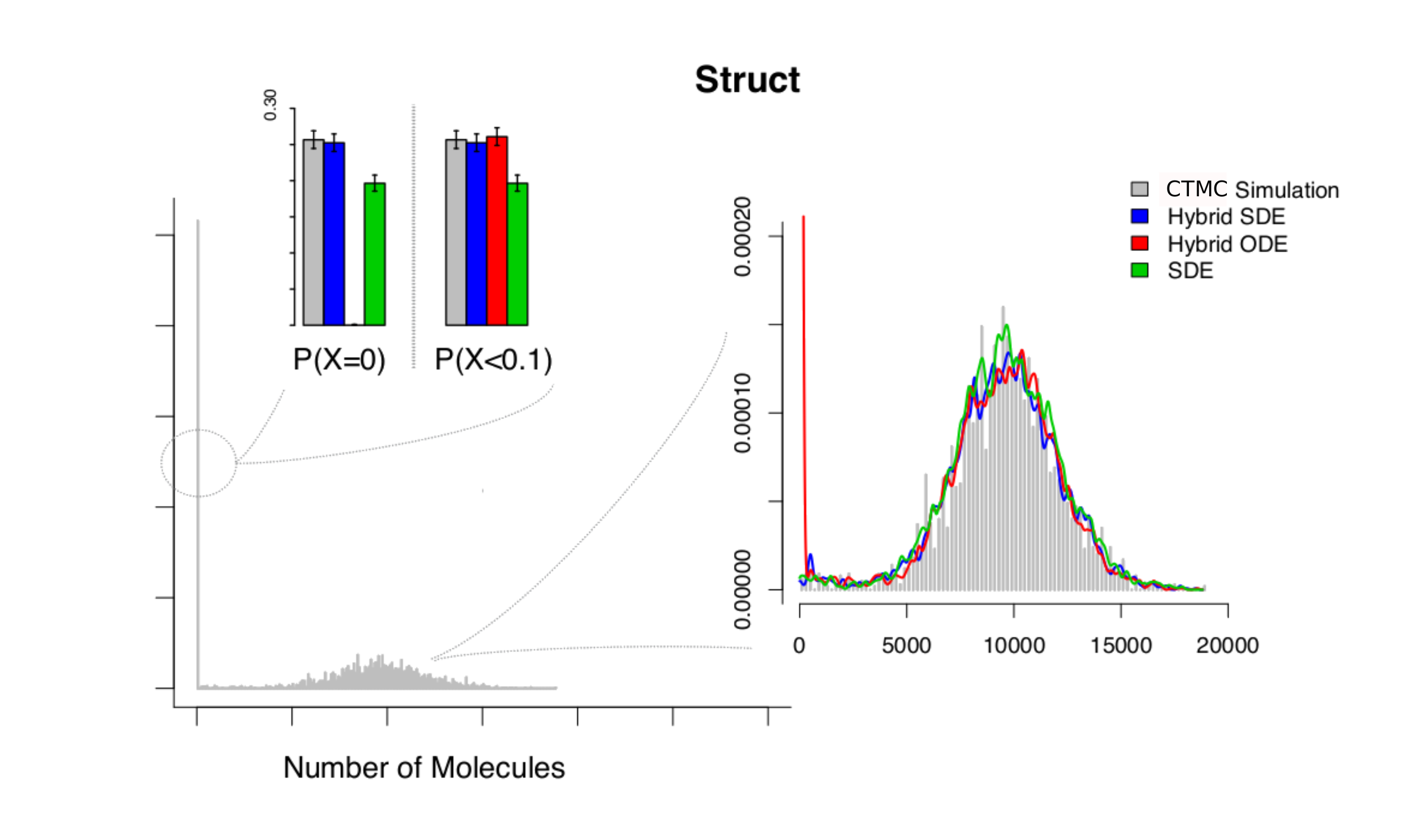}
\caption{Distribution of \emph{struct} moleculas at time 200.}\label{Fig:2}
\end{figure}

A better picture of how the four approximations work is provided by Figure
\ref{Fig:2} where we focus on a single time instant to observe the
distribution of the number of molecules of \emph{struct}.  In particular,
we provide the comparison of the probability distribution of \emph{struct}
after 200 days computed by Monte Carlo simulation against those computed by
using SDE, HODE and SDE.  According to the figure, all three approaches
provide a good representation of the original distribution by reproducing
the bi-stability of the original process and the overall profile of the
distribution.  In particular, Figure~\ref{Fig:2} is structured in such a
way that:
\begin{itemize}
\item On the background, we provide the distribution obtained by simulating
  the original CTMC in order to show that it is extremely sparse over the
  interval $[0,20000]$ and characterized by a large probability mass, about
  $25$ per cent, in zero.
\item On the top of the left side of the figure, we focus on the probability to
  observe the quantity of $struct$ smaller than 0.1 and the probability to
  find $struct$ extinct; in particular, we provide the comparison between
  these results obtained by simulating the original process and those
  generated by using SDE, HODE and HSDE.
\item On the right side of the figure, we provide the comparison between the kernel
  estimates of the probability density function obtained by means of SDE,
  HODE and HSDE and the histogram generated by using the result of the
  simulation.
\end{itemize}
Figure \ref{Fig:2} shows that, although all the three fluid approximations
are able to reproduce the shape of the original distribution, they have
very different behaviors around zero.  Specifically, the HSDE approach is
able to provide an accurate estimate of the probability mass that is
present both \emph{on} the barrier (e.g. in zero) and \emph{around} the barrier (less
than 0.1 molecules) whereas: i) the HODE approach, which is by definition
not able to reach the barrier, fails to represent the first measure, but
provides an accurate estimate of the probability mass present in the
interval $(0,0.1)$; ii) the SDE approximation underestimates both
measures.  This underestimation is in agreement with the overestimation of
the expected number of molecules of \emph{struct} depicted in Figure
\ref{Fig:1}.

Even if all the approaches, but the one based on the ODEs, provide a good approximation
of the bi-modal distribution, it is important to highlight that the four
methods have very different computational costs.  By using an Euler's step
of $0.05$, the solution of the ODE system required few milliseconds.
Keeping the same integration step and computing 5000 trajectories, the SDE
approach required $\approx 8$ seconds, the integration of the HODE system has
been obtained in $\approx 15$ seconds and, finally, the computation of the
HSDE trajectories has been carried out in $\approx 24$ seconds.  Thus, the
SDE approximation is computationally the cheapest among those that provide a good estimate of the
distribution of the process.  Finally the Monte Carlo simulation required $\approx 210$
minutes to compute the same number of trajectories.

\subsection{Transcription regulation}

The second model we consider, a transcriptional regulatory system
described in \cite{Goutsias2005}, consists of the following $9$
reactions
\begin{eqnarray}
\notag
mRNA&\stackrel{k_1}{\rightarrow}& mRNA + M\\
\notag
M &\stackrel{k_2}{\rightarrow}& \emptyset \\
\notag
DNA \cdot D&\stackrel{k_3}{\rightarrow}&DNA\cdot D + mRNA \\
\notag
mRNA &\stackrel{k_4}{\rightarrow}& \emptyset \\
\notag
DNA +D	&\stackrel{k_5}{\rightarrow}& DNA\cdot D \\
\notag
DNA\cdot D &\stackrel{k_6}{\rightarrow}& DNA+D\\
\notag
DNA\cdot D +D&\stackrel{k_7}{\rightarrow}& DNA\cdot2D\\
\notag
DNA\cdot 2D &\stackrel{k_8}{\rightarrow}& DNA\cdot D +D\\
\notag
2M &\stackrel{k_9}{\rightarrow}& D\\
\notag
D &\stackrel{k_{10}}{\rightarrow}& 2M\\
\end{eqnarray}
and is graphically represented by the SPN reported in Figure~\ref{Fig:Exp2} without considering the sub-net in the dashed box.

In details, \emph{mRNA} models the messenger RNA (mRNA) which is
translated into a protein $M$  by the reaction $k_1$. The mRNA transcription (i.e reaction $k_3$) can happen
only when the transcription factor $D$ occupies the DNA binding site $R_1$.
Hence, the DNA binding in position
$R_1$ of $D$ is modeled by reaction $k_5$; while its
unbinding is represented by  reaction $k_6$.  Moreover, we assume that a further binding in position
$R_2$, disabling the basal transcription of mRNA, can happen only when the
binding $R_1$ is already occupied by $D$.  This is modeled by reaction $k_7$.
Its corresponding unbinding is instead modeled by the reaction
$k_8$ (i.e. reaction 8).  The dimerization of $M$ and $D$ is represented by
the reactions $k_9$ and $k_{10}$.
Reactions $k_2$ and $k_4$ model the degradation of mRNA and $M$.

In order to highlight that the behavior at the boundary of the state space
of the fluid components may have a strong impact on the overall dynamics of
the system, we extend the model by adding the following three reactions:
\begin{eqnarray}
\notag
E + M  &\stackrel{k_{11}}{\rightarrow}& EM \\
\notag
EM  &\stackrel{k_{12}}{\rightarrow}& E+M \\
\label{eq:michaelisMenten}
EM  &\stackrel{k_{13}}{\rightarrow}& E + P
\end{eqnarray}
where $E$ is an enzyme catalyst of the production of a protein $P$, and
$EM$ is the complex generated by the binding between $E$ and $M$. The
corresponding sub-net is represented in the dashed box of Figure
\ref{Fig:Exp2} in~\ref{App:1}.  These three additional reactions model the conversion of a
protein $M$ into a new protein $P$ catalyzed by an enzyme $E$ according to
the well-known mass-action enzyme kinetics \cite{segel}.  In particular, by
means of reaction $k_{11}$, the enzyme $E$ binds with the protein $M$ to
form the complex $EM$ which in turn is converted into the product $P$ and
the enzyme $E$ through reaction $13$.  Finally, reaction $k_{12}$
represents the unbinding between $E$ and $M$.  We assumed as kinetic
constants those reported in Table \ref{tab:exp1rates}, and as an initial
state, 40 molecules of $D$, 2 molecules of \emph{DNA}, and $80$ of $E$ (Table\ref{tab:exp2initiL}).
From the structural analysis of this set of reactions we can observe that $DNA$,
$DNA\cdot D$ and $DNA\cdot 2 D$ are part of an invariant, so that the sum
of their corresponding molecules is always constant.  In our experiments this
constant is set equal to $2$.  In this situation, also SDE fails because
\emph{DNA},\emph{DNA-D} and \emph{DNA-2D} are not large enough to be
approximated with a diffusion process.  For this reason, we considered
\emph{DNA},\emph{DNA-D} and \emph{DNA-2D} as discrete quantities and performed
the transient analysis of the model up to $720$ seconds by using only the
hybrid approaches.

\begin{table}[!h]
\footnotesize
\centering
\begin{tabular}{|c c c c c c c c c c c c|}
\hline
$\mu_1$ & $\mu_2$ &$\mu_3$ &$\mu_4$ &$\mu_5/N$ & $\mu_6$ & $\mu_7/N$ & $\mu_8$ & $\mu_9/(2\cdot N)$ & $\mu_{10}$ & $\mu_{11}/N$ &
$\mu_{12}$ \\
\hline
$0.043$ & $0.0001$ & $0.72$ & $0.0039$ & $0.014$ & $0.48$ & $0.00014$ &$8.8~10^{-12}$  & $0.029$ & $0.5$ & $0.001$ & $0.0001$ \\
\hline
\end{tabular}
\caption{Rates of the virus model in $sec^{-1}$}\label{tab:exp2rates}
\end{table}

\begin{table}[!h]
\footnotesize
\centering
\begin{tabular}{|c c c c c c c c c c|}
\hline
$D$ & \emph{DNA} &\emph{DNA-D} &\emph{DNA-2D} &\emph{mRNA} &$M$ & $E$ & $EM$ & $P$ & $N$\\
\hline
$40$ & $2$ & $0$ & $0$ & $0$ & $0$ & $80$ &$0$  & $0$  & 650\\
\hline
\end{tabular}
\caption{Initial condition of the virus model}\label{tab:exp2initiL}
\end{table}
%
%
%



\begin{figure}[!h]
 \includegraphics[width=\textwidth]{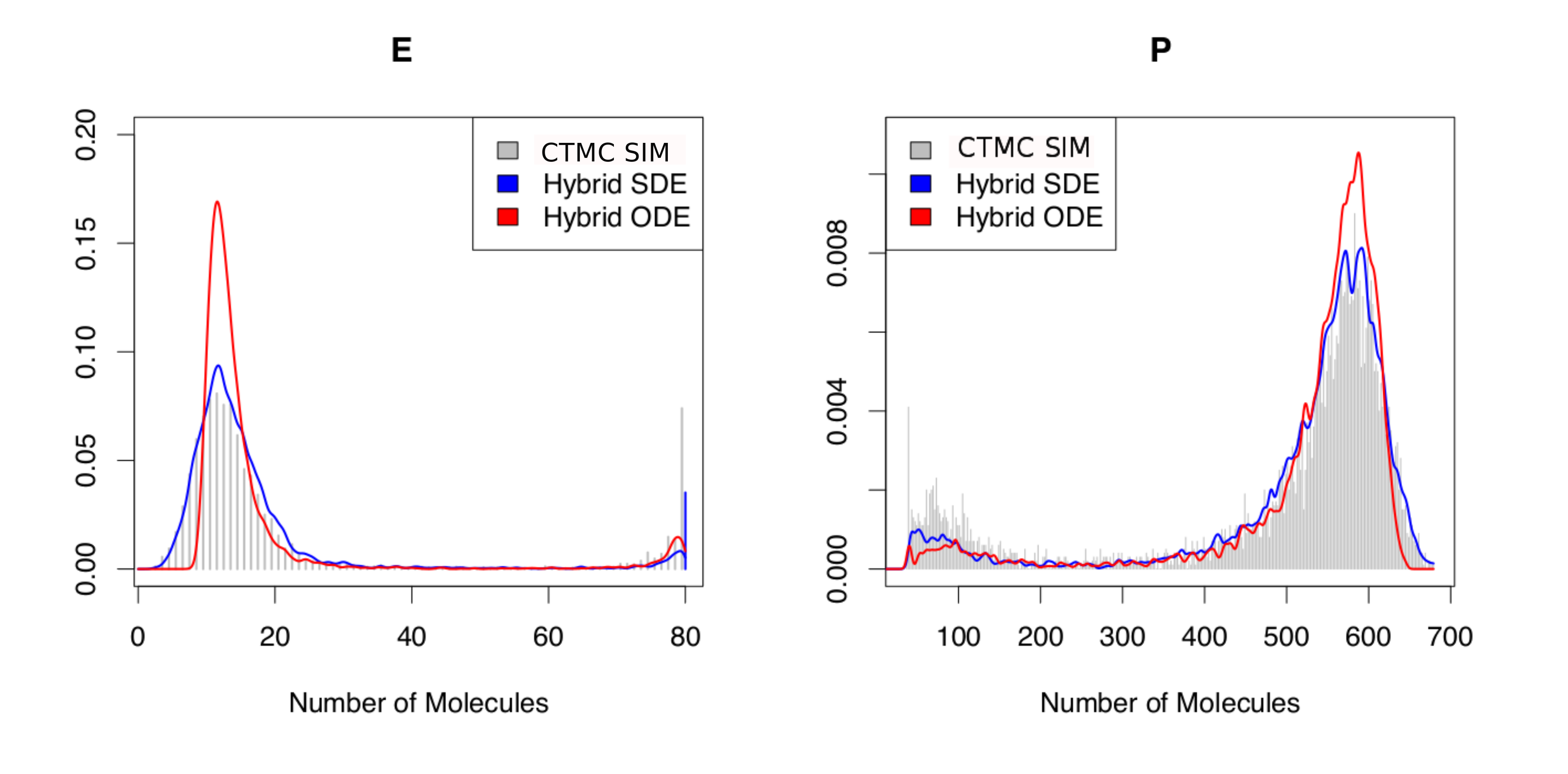}
\caption{Comparison between distributions computed with Monte Carlo Simulation, HSDE, HODE.}\vspace{-0.5cm}\label{Fig:3}
\end{figure}

Figure \ref{Fig:3}(a) provides the comparison between the probability
distributions of \emph{E} after 720 seconds obtained with Monte Carlo
simulation and those computed with the HSDE and the HODE approaches whereas
Figure \ref{Fig:3}(b) reports the same measures obtained for \emph{M}. 
Both the figures highlight how the HSDE approach provides a better approximation with respect to that of the HODE.
Indeed, the HODE approach shows (in both the figures)  a peak that is almost the double of that provided by the
simulation of the original process.  Furthermore, HSDE gives a good
estimate of the probability mass present on the barriers whereas by
construction the HODE approach is unable to reach the barriers.  At last, 
the computation of 5000 trajectories by using 0.05 as integration step
required $\approx 30$ seconds with the HODE approach and $\approx 74$ with
HSDE.  The Monte Carlo simulation of the same number of trajectories
required $\approx 10$ minutes.

\subsection{The tool}\label{tool}
\begin{figure}[tbp]
   \centering
   \includegraphics[width=0.70\textwidth]{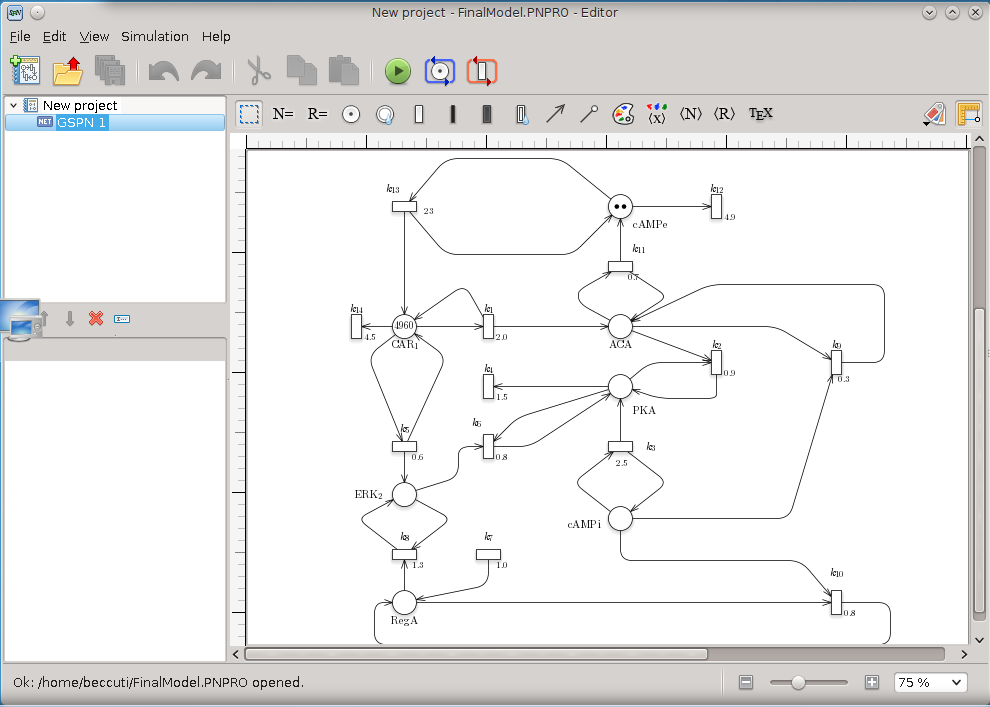}
   \caption{GreatSPN GUI.}\label{Fig:Arc1}
\end{figure}

The comparison of the different approaches has been carried out with a
prototype implementation integrated in the GreatSPN framework
\footnote{A VirtualBox image, in which this prototype implementation is installed, is available on demand sending an email to greatspn@di.unito.it.}~\cite{BabarBDM10}.
This allows us to use the new GreatSPN GUI (see Fig.~\ref{Fig:Arc1}) to easily design
the SPN model and to generate  the corresponding (H)ODE/(H)SDE solver.

\begin{figure}[tbp]
   \centering
   \includegraphics[width=0.60\textwidth]{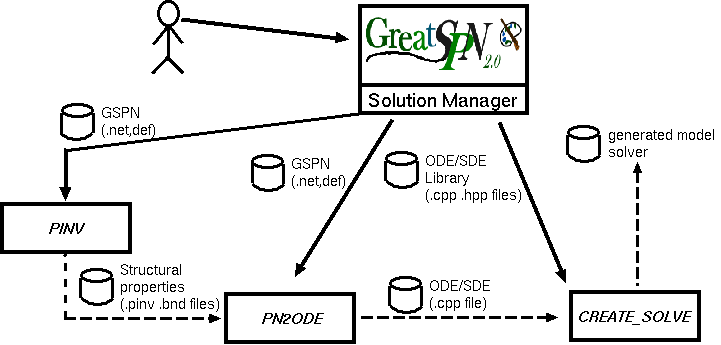}
   \caption{Framework architecture.}\label{Fig:Arc}
\end{figure}
This generation process requires the following three steps:
\begin{enumerate}
\item \emph{PINV} computes the (lower and upper) bounds  for all the quantities involved in the systems.
\item \emph{PN2ODE} generates from an SPN model a C++ file implementing the corresponding (H)ODE/(H)SDE system. This system is directly encoded in C++ exploiting a set of specific classes (e.g. \emph{SystemEquation}, \emph{Equation}, \ldots) defined in an ad-hoc developed library;
\item \emph{CREATE\_SOLVE}
compiles the previously generated C++ code with
  the ad-hoc developed library to generate the corresponding (H)ODE/(H)SDE solver.
In particular this library implements the standard Euler method to solve ODE systems, the Euler-Maruyama method~\cite{klebaner} to solve SDE systems\footnote{Currently, we are working to implement Runge–Kutta method in our framework, however it is important to highlight that the use of methods with better convergence order does not change the quality of the results presented in this section.}, and  the First-Reaction method~\cite{pahle} to simulate them.
Moreover, it provides an extension of  Euler-Maruyama method  for computing the  numerical solution of our  (hybrid) jump diffusion approximation (see \ref{app:algo} where the pseudo-code which describes such extension is reported).
\end{enumerate}
Then, the generated solver  can be executed by command line as follows:
\begin{equation*}
{\bf\emph{\emph{net\_name.solver out\_file\_name type  step num\_runs max\_time [-B place\_bounds]}}}
\end{equation*}
where \emph{type} is used to specify which method will be used to solve the system (i.e. ODE, HODE, SDE, HSDE and SIM),
\emph{step} is the maximum Euler step, \emph{num\_runs} is the maximum number of runs (i.e., MaxRuns), and \emph{max\_time} is the final time  for which the solution is computed.
Optionally the user can specify the bounds (lower and upper) for each net place in a separate file, so that she/he will override those automatically computed from the SPN model by PINV (i.e. \emph{-B place\_bounds}).


\section{Conclusions}\label{Sec:con}
In this paper we have provided numerical evidence that Kurtz's diffusion
approximation can be extended to a jump diffusion approximation to address
the case when the process reaches the boundary with non-negligible
probability and that such jump diffusion approximation can be extended
further in the framework of hybrid models.  Our proposal allows to handle
cases in which the number of certain objects represented in the original
model does not grow unbounded, thus violating one of the conditions for
density dependent Markov chains.  In these cases we have shown that it is
possible to apply the jump diffusion approximation only to those components
of the model that are in density dependent form and are associated with
high population levels.  The remaining components are treated as discrete
quantities.  The resulting process is a hybrid switching jump diffusion,
i.e., a Markov process with hybrid state space and jumps where the discrete
state changes can be seen as switches that take the diffusion from one mode
to another.  We have shown that the stochastic differential equations that
characterize this process can be derived automatically both from the
description of the original Markov chains as well as from a model specified
using a high level description language, like stochastic Petri nets.  To
support our proposal, we have applied the method for the analysis of four
models of biological interest: a model of the crazy clock reaction and its
variation with a switching behavior, a model describing viral infection
kinetics and another representing a transcriptional regulatory mechanism.
The results have been obtained in part by numerical integration of the
Fokker Plank equation and in part by Monte Carlo simulations of the new
approximating model.  For all four examples, our method closely reproduces
the behavior of the original CTMC with a substantial saving of execution
time.





\bibliographystyle{elsarticle-num}
\bibliography{bibl}

\begin{thebibliography}{10}
\expandafter\ifx\csname url\endcsname\relax
  \def\url#1{\texttt{#1}}\fi
\expandafter\ifx\csname urlprefix\endcsname\relax\def\urlprefix{URL }\fi
\expandafter\ifx\csname href\endcsname\relax
  \def\href#1#2{#2} \def\path#1{#1}\fi

\bibitem{TuScBu04}
T.~Turner, S.~Schnell, K.~Burrage, Stochastic approaches for modelling in vivo
  reactions, Comp. Bio. Chem. 28 (2004) 165–178.

\bibitem{Wi06}
D.~Wilkinson, Stochastic Modelling for Systems Biology, Chapman \& Hall, 2006.

\bibitem{Gillespie_1977}
D.~T. Gillespie, {Exact stochastic simulation of coupled chemical reactions},
  J. Phys. Chem. 81~(25) (1977) 2340--2361.

\bibitem{Ku70}
T.~G. Kurtz, Solutions of ordinary differential equations as limits of pure
  jump {M}arkov processes, Journal of Applied Probability 1~(7) (1970) 49--58.

\bibitem{St95}
W.~J. Stewart, Introduction to the Numerical Solution of {M}arkov Chains.,
  Princeton University Press, Princeton, New Jersey, USA, 1995.

\bibitem{Gi01}
D.~T. Gillespie, {Approximate accelerated stochastic simulation of chemically
  reacting systems}, J Chem Phys 115 (2001) 1716--1733.

\bibitem{RaPeCaGi03}
M.~Rathinam, L.~R. Petzold, Y.~Cao, D.~T. Gillespie, {Stiffness in stochastic
  chemically reacting systems: {T}he implicit tau-leaping method}, J Chem Phys
  119~(24) (2003) 12784--12794.

\bibitem{CaGiPe05}
Y.~Cao, D.~T. Gillespie, L.~R. Petzold, {The slow-scale stochastic simulation
  algorithm}, J Chem Phys 122~(1).

\bibitem{DaMiWo10}
T.~Dayar, L.~Mikeev, V.~Wolf, On the numerical analysis of stochastic
  {L}otka-{V}olterra models, in: Proc. of the Workshop on Computer Aspects of
  Numerical Algorithms (CANA�10), 2010, pp. 289--296.

\bibitem{MaWoDiHe10}
M.~Mateescu, V.~Wolf, F.~Didier, T.~A. Henzinger, Fast adaptive uniformisation
  of the chemical master equation, IET systems biology 4~(6) (2010) 441--452.

\bibitem{ZhWaCa10}
J.~Zhang, L.~T. Watson, Y.~Cao, A modified uniformization method for the
  solution of the chemical master equation, Computers \& Mathematics with
  Applications 59~(1) (2010) 573 -- 584.

\bibitem{ZhWaCa09}
J.~Zhang, L.~T. Watson, Y.~Cao, Adaptive aggregation method for the chemical
  master equation, Int. J. of Computational Biology and Drug Design 2~(2)
  (2009) 134--148.

\bibitem{CiDeHiCa09}
F.~Ciocchetta, A.~Degasperi, J.~Hillston, M.~Calder, Some investigations
  concerning the {CTMC} and the ode model derived from bio-pepa., Electron.
  Notes Theor. Comput. Sci. 229 (2009) 145--163.

\bibitem{CoCoCo11}
F.~Cordero, A.~Horv{\'a}th, D.~Manini, L.~Napione, M.~D. Pierro, S.~Pavan,
  A.~Picco, A.~Veglio, M.~Sereno, F.~Bussolino, G.~Balbo, Simplification of a
  complex signal transduction model using invariants and flow equivalent
  servers, Theor. Comput. Sci. 412~(43) (2011) 6036--6057.

\bibitem{AnHo11b}
A.~Angius, A.~Horv\'ath, Product form approximation of transient probabilities
  in stochastic reaction networks, Electronic Notes on Theoretical Computer
  Science 277 (2011) 3--14.

\bibitem{AAAHVW}
A.~Angius, A.~Horv{\'a}th, V.~Wolf, Quasi product form approximation for markov
  models of reaction networks, Transactions on Computational Systems Biology
  7625~(XIV).

\bibitem{kurtz1976limit}
T.~G. Kurtz, Limit theorems and diffusion approximations for density dependent
  markov chains, in: Stochastic Systems: Modeling, Identification and
  Optimization, I, Springer, 1976, pp. 67--78.

\bibitem{andersonKurtz}
D.~F. Anderson, T.~G. Kurtz, Continuous time markov chain models for chemical
  reaction networks, in: Design and Analysis of Biomolecular Circuits,
  Springer, 2011, pp. 3--42.

\bibitem{ATPN14}
M.~Beccuti, E.~Bibbona, A.~Horv\'ath, R.~Sirovich, A.~Angius, G.~Balbo,
  Analysis of {P}etri net models through stochastic differential equation, in:
  Proc. of International Conference on Application and Theory of Petri Nets and
  other models of concurrency (ICATPN'14), Tunis, Tunisia, 2014,
  \href{http://arxiv.org/abs/1404.0975}{arXiv:1404.0975}.

\bibitem{yin}
G.~G. Yin, C.~Zhu, \href{http://dx.doi.org/10.1007/978-1-4419-1105-6}{Hybrid
  switching diffusions}, Vol.~63 of Stochastic Modelling and Applied
  Probability, Springer, New York, 2010, properties and applications.
\newblock \href {http://dx.doi.org/10.1007/978-1-4419-1105-6}
  {\path{doi:10.1007/978-1-4419-1105-6}}.
\newline\urlprefix\url{http://dx.doi.org/10.1007/978-1-4419-1105-6}

\bibitem{HoKuNiTr98}
G.~Horton, V.~G. Kulkarni, D.~M. Nicol, K.~S. Trivedi, Fluid stochastic petri
  nets: Theory, application, and solution techniques, Eur. J. Op. Res. 105~(1)
  (1998) 184--201.

\bibitem{GrSeHoBo01}
M.~Gribaudo, M.~Sereno, A.~Horv\'ath, A.~Bobbio, Fluid stochastic {P}etri nets
  augmented with flush-out arcs: Modelling and analysis, Discrete Event Dynamic
  Systems: Theory and Applications 11 (2001) 97--117.

\bibitem{PSP:2044020}
D.~R. Cox, The analysis of non-markovian stochastic processes by the inclusion
  of supplementary variables, Mathematical Proceedings of the Cambridge
  Philosophical Society 51 (1955) 433--441.

\bibitem{JANE}
A.~Pourranjbar, J.~Hillston, L.~Bortolussi, Don{\textquoteright}t just go with
  the flow: Cautionary tales of fluid flow approximation, in: M.~Tribastone,
  S.~Gilmore (Eds.), EPEW 2012, and UKPEW 2012, Vol. 7587 of Lecture Notes in
  Computer Science, Springer, Springer, 2012, p. 156{\textendash}171.

\bibitem{kurtz1978strong}
T.~G. Kurtz, Strong approximation theorems for density dependent markov chains,
  Stochastic Processes and Their Applications 6~(3) (1978) 223--240.

\bibitem{klebaner}
F.~C. Klebaner, Introduction to stochastic calculus with applications, 3rd
  Edition, Imperial College Press, London, 2012.

\bibitem{rogers}
L.~C.~G. Rogers, D.~Williams, Diffusions, Markov Processes, and Martingales:
  Volume 2, It$\bar{\mbox{o}}$ calculus., Cambridge university press, 2000.

\bibitem{gillespie}
D.~T. Gillespie, The chemical langevin equation, J. Chem. Phys. 113 (2000) 297.

\bibitem{srivastava}
R.~Srivastava, L.~You, J.~Summers, J.~Yin, Stochastic vs. deterministic
  modeling of intracellular viral kinetics, Journal of Theoretical Biology
  218~(3) (2002) 309--321.

\bibitem{BOABCDF95}
M.~Ajmone~Marsan, G.~Balbo, G.~Conte, S.~Donatelli, G.~Franceschinis,
  {Modelling with Generalized Stochastic Petri Nets}, J. Wiley, New York, NY,
  USA, 1995.

\bibitem{pdmp}
M.~H.~A. Davis, Piecewise-deterministic markov processes: A general class of
  non-diffusion stochastic models, Journal of the Royal Statistical Society.
  Series B (Methodological) 46~(3) (1984) pp. 353--388.

\bibitem{hybridLimit}
L.~Bortolussi, Hybrid limits of continuous time markov chains, in: Quantitative
  Evaluation of Systems (QEST), 2011 Eighth International Conference on, 2011,
  pp. 3--12.

\bibitem{pola2003stochastic}
G.~Pola, M.~Bujorianu, J.~Lygeros, M.~Benedetto, Stochastic hybrid models: An
  overview, in: Proc. IFAC Conf. Anal. Design Hybrid Syst, 2003, pp. 45--50.

\bibitem{caravagna}
G.~Caravagna, A.~d'Onofrio, M.~Antoniotti, G.~Mauti, {Stochastic Hybrid
  Automata with delayed transitions to model biochemical systems with delays},
  {INFORMATION AND COMPUTATION} {236}~({SI}) ({2014}) {19--34}.
\newblock \href {http://dx.doi.org/{10.1016/j.ic.2014.01.010}}
  {\path{doi:{10.1016/j.ic.2014.01.010}}}.

\bibitem{menz}
S.~Menz, J.~C. Latorre, C.~Schütte, W.~Huisinga, Hybrid
  stochastic--deterministic solution of the chemical master equation,
  Multiscale Modeling \& Simulation 10~(4) (2012) 1232--1262.

\bibitem{verena}
T.~A. Henzinger, L.~Mikeev, M.~Mateescu, V.~Wolf, Hybrid numerical solution of
  the chemical master equation, in: Proceedings of the 8th International
  Conference on Computational Methods in Systems Biology, ACM, 2010, pp.
  55--65.

\bibitem{luca2010}
L.~Bortolussi, Limit behavior of the hybrid approximation of stochastic process
  algebras, in: K.~Al-Begain, D.~Fiems, W.~Knottenbelt (Eds.), 17th
  International Conference on Analytical and Stochastic Modeling Techniques and
  Applications (ASMTA 2010), Springer-Verlag, Springer-Verlag, 2010, p.
  367{\textendash}381.

\bibitem{intep}
S.~Intep, D.~J. Higham, X.~Mao, Switching and diffusion models for gene
  regulation networks, Multiscale Modeling \& Simulation 8~(1) (2009) 30--45.

\bibitem{haseltine}
E.~L. Haseltine, J.~B. Rawlings, Approximate simulation of coupled fast and
  slow reactions for stochastic chemical kinetics, The Journal of chemical
  physics 117~(15) (2002) 6959--6969.

\bibitem{salis}
H.~Salis, Y.~Kaznessis, Accurate hybrid stochastic simulation of a system of
  coupled chemical or biochemical reactions, The Journal of chemical physics
  122~(5) (2005) 054103.

\bibitem{gshs}
M.~Bujorianu, J.~Lygeros, Toward a general theory of stochastic hybrid systems,
  in: H.~Blom, J.~Lygeros (Eds.), Stochastic Hybrid Systems, Vol. 337 of
  Lecture Notes in Control and Information Science, Springer Berlin Heidelberg,
  2006, pp. 3--30.

\bibitem{lyg1}
J.~Hu, J.~Lygeros, S.~Sastry, Towards a theory of stochastic hybrid systems,
  in: N.~Lynch, B.~Krogh (Eds.), Hybrid Systems: Computation and Control, Vol.
  1790 of Lecture Notes in Computer Science, Springer Berlin Heidelberg, 2000,
  pp. 160--173.

\bibitem{griffith}
M.~Griffith, T.~Courtney, J.~Peccoud, W.~Sanders, Dynamic partitioning for
  hybrid simulation of the bistable {HIV}-1 transactivation network,
  Bioinformatics 22~(22) (2006) 2782--2789.

\bibitem{salis2}
H.~Salis, V.~Sotiropoulos, Y.~Kaznessis, Multiscale hy3s: Hybrid stochastic
  simulation for supercomputers, BMC Bioinformatics 7.

\bibitem{feller1d}
W.~Feller, Diffusion processes in one dimension, Trans. Amer. Math. Soc. 77
  (1954) 1--31.

\bibitem{fellerParabolic}
W.~Feller, The parabolic differential equations and the associated semi-groups
  of transformations, Ann. of Math. (2) 55 (1952) 468--519.

\bibitem{ErdiLente}
P.~{\'E}rdi, G.~Lente, Stochastic Chemical Kinetics, Springer Series in
  Synergetics, Springer, 2014.

\bibitem{BabarBDM10}
J.~Babar, M.~Beccuti, S.~Donatelli, A.~S. Miner, Greatspn enhanced with
  decision diagram data structures, in: Proceedings of Applications and Theory
  of Petri Nets, 31st International Conference, PETRI NETS 2010, Braga,
  Portugal, June 21-25,, IEEE Computer Society, Los Alamitos, California, USA,
  2010, pp. 308--317.

\bibitem{Rurl}
{{R} webpage}, \url{http://www.r-project.org/}.

\bibitem{Goutsias2005}
J.~Goutsias, {Quasiequilibrium approximation of fast reaction kinetics in
  stochastic biochemical systems}, The Journal of Chemical Physics
  122~(184102).
\newblock \href {http://dx.doi.org/10.1063/1.1889434}
  {\path{doi:10.1063/1.1889434}}.

\bibitem{segel}
I.~Segel, \href{http://books.google.it/books?id=6uFqAAAAMAAJ}{Enzyme kinetics:
  behavior and analysis of rapid equilibrium and steady state enzyme systems},
  Wiley classics library, Wiley, 1975.
\newline\urlprefix\url{http://books.google.it/books?id=6uFqAAAAMAAJ}

\bibitem{pahle}
J.~Pahle, Biochemical simulations: stochastic, approximate stochastic and
  hybrid approaches, Briefings in Bioinformatics 10(1) (2009) 53--64.

\bibitem{bremaud1999}
P.~Br{\'e}maud, Markov chains, Vol.~31 of Texts in Applied Mathematics,
  Springer-Verlag, New York, 1999, gibbs fields, Monte Carlo simulation, and
  queues.

\bibitem{molloy:spn}
M.~K. Molloy, Performance analysis using stochastic petri nets, IEEE
  Transactions on Computers 31~(9) (1982) 913--917.
\newblock \href {http://dx.doi.org/http://dx.doi.org/10.1109/TC.1982.1676110}
  {\path{doi:http://dx.doi.org/10.1109/TC.1982.1676110}}.

\end{thebibliography}

\appendix

\section{Examples and properties of density dependent processes\label{app:dd}}

\subsection{Example for a density dependent family\label{app:ddex}}

We describe here a family of density dependent CTMCs which corresponds to a
simple epidemic model.  The groups involved in the process are that of the
susceptible individuals and that of the infected ones.  The state of the
process is a pair $(i,j)$ giving the number of susceptible individuals,
$i$, and the number infected ones, $j$.  We consider an area of size $N$
and assume that there are three possible events.  The transition that makes
the number of susceptible individuals grow is proportional to the size of
the area, i.e.,
\[
q_{(i,j),(i+1,j)}=N \lambda_1
\]
Infection occurs by the interaction of a susceptible individual and an
already infected one.  The intensity of the corresponding reaction in state
$(i,j)$ is proportional to the product $i \times j$ and inversely
proportional to the size of the considered area, i.e., we have
\[
q_{(i,j),(i-1,j+1)}=\lambda_2 \frac{ij}{N}
\]
Infected individuals are assumed to become immune independently of each
other and independently of the size of the area, i.e., 
\[
q_{(i,j),(i,j-1)}=\lambda_3 j
\]
All the above three intensities can be written in such a form that they are
proportional to the size of the area and they depend on the density of the
number of the individuals, $i/N$ and $j/N$, (or do not depend on one or
both state variables)
\[
q_{(i,j),(i+1,j)}=N \lambda_1,~~~~q_{(i,j),(i-1,j+1)}=N\lambda_2 \left(\frac{i}{N}\right)\left(\frac{j}{N}\right),~~~~
q_{(i,j),(i,j-1)}=N \lambda_3 \left(\frac{j}{N}\right)
\]
Density dependent CTMCs are those whose transition intensities are in the
above form.  The formal definition is provided in
Definition~\ref{def:density_dependent}.

For the above model the set of possible state changes is
$C=\{(1,0),(-1,+1),(0,-1)\}$ and function $f$ required by
Definition~\ref{def:density_dependent} is defined as
\[
f(y,l)=\left\{
\begin{array}{ll}
 \lambda_1 &\mbox{if}~l=(+1,0) \\
 \lambda_2 y_1 y_2 &\mbox{if}~l=(-1,+1) \\
 \lambda_3 y_2 &\mbox{if}~l=(0,-1) \\
 0 & \mbox{otherwise}
\end{array}
\right.
\]

\subsection{Properties of density dependent processes\label{app:prop}}

In order to gain a better understanding of the property of density
dependence, let us introduce some general concept from the theory of Markov
chains. Among the many books devoted to this topic, we refer the reader
to \cite{bremaud1999}.

For a general Markov chain $M(u)$ with state space $S \in \mathbb{Z}^{k}$
and instantaneous transition rates $q_{i,j}$, let us introduce the
following key object
\begin{align}\label{eq:F}
F_{M}(i) = \sum_{j\in S} (j-i) q_{i,j} \quad i \in S.
\end{align}
The function $F_{M}$ will be referred to as the \emph{generator} of the
chain.  Under suitable hypothesis the expectation of $M(u)$ solves the
following \emph{Dynkin equation} cf. \cite[Chapter 9, Theorem
2.2]{bremaud1999}
\begin{align}\label{eq:Dynkin}
\frac{d \mathbb{E}(M(u))}{du} = \mathbb{E}[F_{M}(M(u))].
\end{align}

Two invariance properties of density dependent CTMCs can be then stated
using the normalized chains.

\begin{property}\label{rmk:genC}
The density dependence property of the family $X\BN(u)$ is equivalent to
require that for the family of the normalized CTMCs, $Z\BN(u)$, the
generator is constant.  This constant generator, denoted by $F(y)$, is
equal to $F_{Z\BN}(y)$ for all $N$ and hence
\begin{align}
\label{eq:gen_cost}
F(y) = \sum_{l \in C} \frac{l}{N} \cdot
p \BN _{y,y+\frac{l}{N}} = \sum_{l \in C} \frac{l}{N} \cdot
q \BN _{Ny,Ny+l} = \sum_{l\in C} l f\left(y,l\right)
\end{align}
where $p \BN _{i,j}$ and $q \BN _{i,j}$ are the instantaneous transition
rates of the processes $Z \BN $ and $X \BN $, respectively.  Note that a
necessary condition for a generator to be constant is that the entries of
the vectors that describe the effect of the transitions, i.e., the entries
of the vectors in $C$, cannot depend on $N$.
\end{property}

\begin{property}
Each element of the family $Z \BN (u)$ solves the same Dynkin equation
\begin{align}\label{eq:dynkinZ}
\frac{d \mathbb{E}(Z \BN (u))}{du} = \mathbb{E}[F(Z \BN (u))].
\end{align}
Note that (\ref{eq:odeK}) is analogous to the above Dynkin equation.  
\end{property}

In case of a nearly density dependent family Property~\ref{rmk:genC} does
not hold but, as stated by the following property, the generator is still
dominated by the function $F(y)$ given in (\ref{eq:gen_cost}) as the
indexing parameter increases.
\begin{property}
If the family $X \BN (u)$ is nearly density dependent
then 
\[
F_{Z \BN }\left(y \right) = F(y)+
 O\left( \frac{1}{N} \right),
\]
from which it follows that
\begin{equation}
\label{eq:gen_lim} 
\lim_{N \rightarrow \infty}F_{Z \BN }\left(y \right) = F(y)
\end{equation}
\end{property}

\subsection{Petri Nets\label{sec:pn}}

Petri Nets (PNs) are bipartite directed graphs with two types of nodes:
places and transitions.  The places, graphically represented as circles,
correspond to the state variables of the system (e.g., chemical compounds),
while the transitions, graphically represented as rectangles, correspond to
the events (e.g., reaction occurrences) that can induce state changes.  The
arcs connecting places to transitions (and vice versa) express the
relations between states and event occurrences.  Places can contain tokens
(e.g., molecules) drawn as black dots within the places. The state of a PN,
called marking, is defined by the number of tokens in each place.  The
evolution of the system is given by the occurrence of enabled transitions,
where a transition is enabled iff each of its input places contains a number
of tokens greater than or equal to a given threshold defined by the
cardinality of the corresponding input arc.  A transition occurrence,
called firing, removes a fixed number of tokens from its input places and
adds a fixed number of tokens to its output places (according to the
cardinality of its input/output arcs).

Stochastic Petri Nets (SPNs) are PNs where the firing of each transition is
assumed to occur after a random delay (firing time) from the time it is
enabled.  In this paper we consider exponentially distributed random delays
\cite{molloy:spn}.  Accordingly, each transition is associated with a
rate that represents the parameter of its firing delay distribution.
Firing rates may be marking dependent.  When a marking is entered an
exponentially distributed random delay is sampled for each enabled
transition.  The transition with the lowest delay fires and the system
changes marking accordingly.  Consequently, the underlying stochastic
process is a CTMC.

Formally, an SPN is defined by the following ingredients.  $P$ and
$T$ are the sets of places and transitions, respectively.  The initial
marking is given by a vector of non-negative integers of length $|P|$ and
it is denoted by ${\bf m_0}$.  The multiplicity of the input arcs of
transition $t \in T$ is given by a vector of non-negative integers of
length $|P|$ and it is denoted by $I_{t}$.  Similarly, the multiplicity of
the output arcs are given by the vector denoted by $O_{t}$.  

\begin{figure}[tbp]
   \centering
   \includegraphics[width=0.40\textwidth]{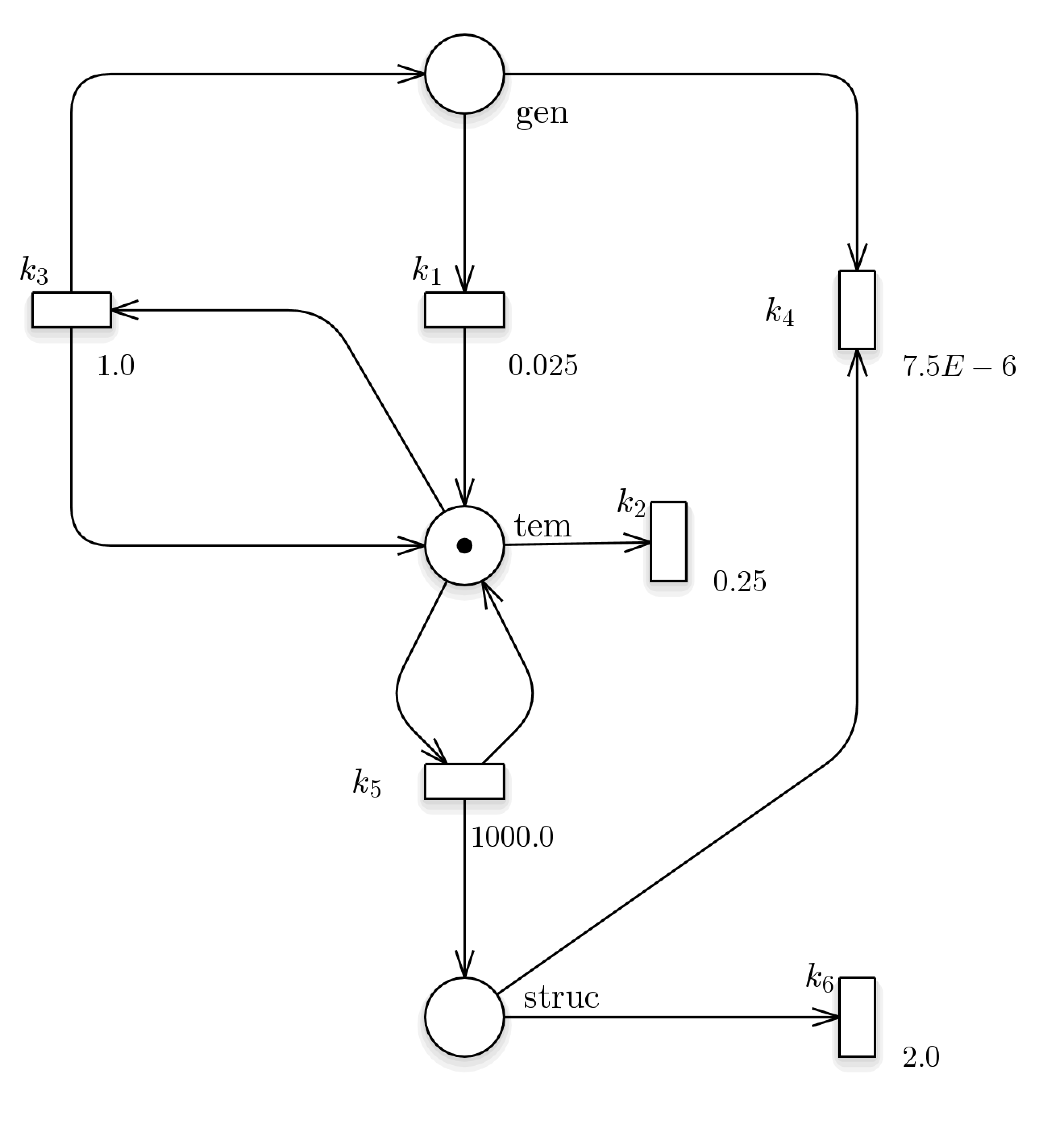}
   \caption{PN model describing the intracellular kinetics of a generic virus}\label{Fig:Exp1}
\end{figure}

\begin{figure}[tbp]
   \centering
   \includegraphics[width=0.7\textwidth]{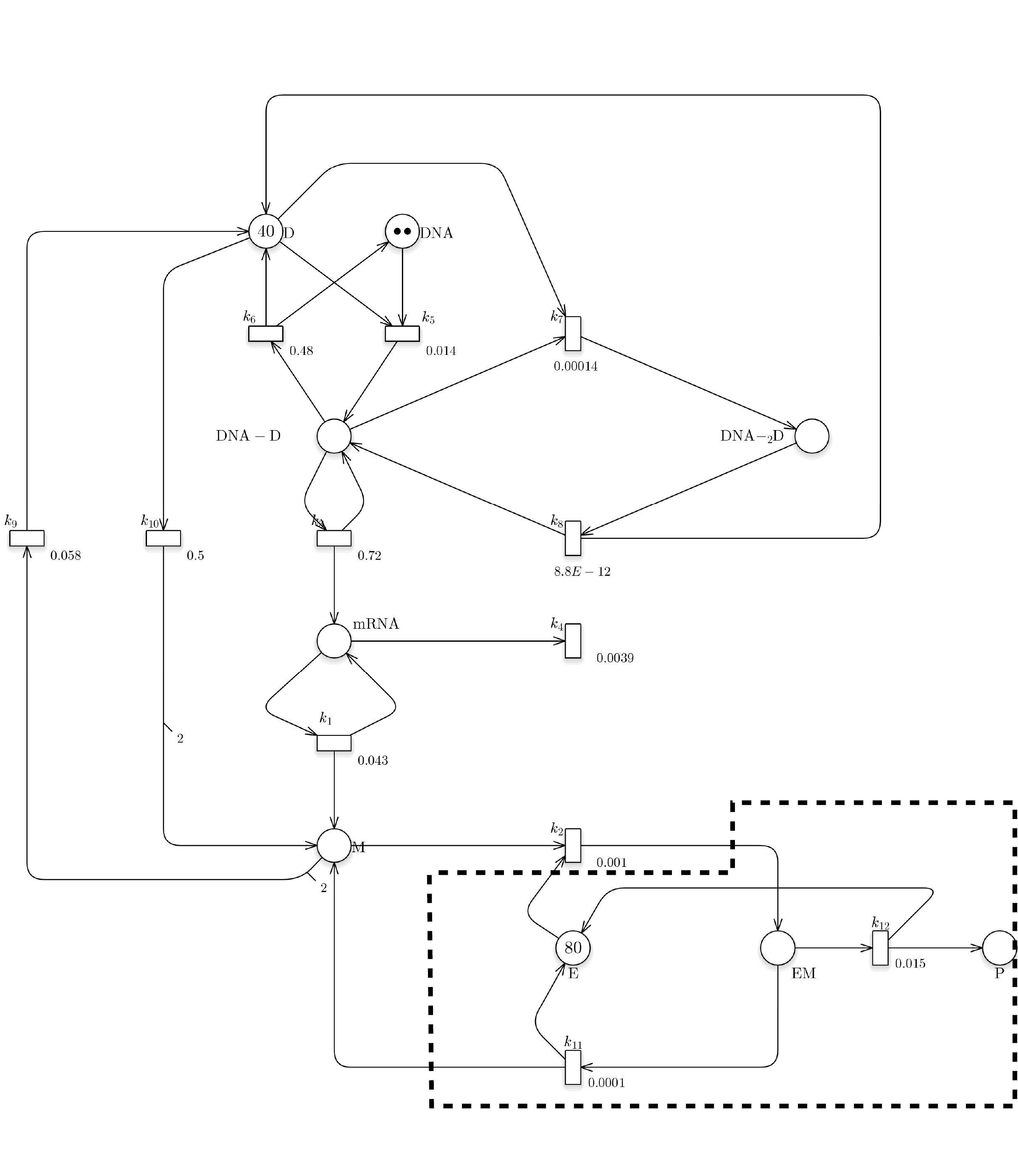}
   \caption{SPN model describing the transcription regulation.}\label{Fig:Exp2}
\end{figure}

In Figs.\ref{Fig:Exp1} and\ref{Fig:Exp2} are shown the  SPN models for the two case studies presented in Sec.\ref{Sec:res}.

The parameter
of the exponential distribution of the firing delay of transition $t$ in
marking ${\bf m}$ is denoted by $\lambda_{t,{\bf m}}$.
With the above notation the set of possible state changes is given by
$C=\{l~|~l=O_t-I_t, t \in T\}$ because the overall effect of transition $t$ is
given by $O_t-I_t$.  Transition $t$ is enabled in marking ${\bf m}$ iff ${\bf
m} \geq I_t$.  And the transition intensities of the CTMC corresponding to
the SPN can be written as
\[
q_{{\bf m},{\bf m}+l}=
\sum_{\forall t: l=O_t-I_t \wedge {\bf m} \geq I_t}
\lambda_{t,{\bf m}}
\]

A special form of transition intensity, which is particularly interesting
in our context, arises when a transition models a chemical reaction
behaving according to the stochastic law of mass action.  In this case
the intensity of transition $t$ in marking ${\bf m}$ is given as
\begin{equation}
\label{eq:lme}
\lambda_{t,{\bf m}}=
N^{1-\sum_{i=1}^{|P|} I_t(i)}
\mu_t \prod_{i=1}^{|P|} \binom{{\bf m}(i)}{I_t(i)}
\end{equation}
where $V$ is the volume in which the reaction occurs, $\mu_t$ is the rate
constant of the reaction, and $v(i)$ is the $i$th entry of the vector $v$.
The above can be written separating the contributions with different orders in $N$ as 
\[
\lambda_{t,{\bf m}}=
N \left(
\mu_t
\prod_{i=1}^{|P|} \frac{1}{I_t(i)!} \left(\frac{{\bf m}(i)}{N}\right)^{I_t(i)}
+O\left( \frac{1}{N} \right)
\right)
\]
which shows that models in which all transitions act according to the
stochastic law of mass actions are nearly density dependent and that the
function $f$ required by the definitions of density dependence is given
by
\[
f(y,l)= \sum_{\forall t: l=O_t-I_t \wedge {\bf m} \geq I_t}
\mu_t
\prod_{i=1}^{|P|} \frac{y(i)^{I_t(i)}}{I_t(i)!}
\]

\label{App:1}

\section{Pseudo-code implementing (H)SDE solver\label{app:algo}}

Algorithm~\ref{algo1} shows the pseudo-code implementing a simple
simulation approach to analyze HSJD models in which its quantities can
visit their barrier. The algorithm takes in input a HSJD system
(called \emph{SSDE}), a step size used for the Euler scheme (i.e \emph{step}),
the number of runs (i.e. \emph{MaxRuns}), and a final time
(i.e. \emph{FinalTime}); and it prints the generates traces.

In details, the method \emph{init()} at line~\ref{line1} initializes the
vector \emph{Value} encoding the state of the HSJD; the
method \emph{Copy()}, called at line~\ref{line2} before computing the state
at the next step, copies the current values of vector \emph{Value} in the
vector \emph{PrValue} storing the previous state.  Method
\emph{CheckBound()} taking in input the current state returns the
list of events currently in $\stella{C}^{F}_{x}$.  Among these events the
method \emph{CheckFire()} at line~\ref{line3} returns the first that is
scheduled to fire (i.e., \emph{e}) with firing delay smaller than the
current solution step (i.e., \emph{h}). Moreover, such method re-sets $h$
value according to the selected firing delay time.

Then, for each SDE equation the method \emph{getEvents()} returns the list
of events (i.e. \emph{ListE}) in $\anello{C}^{F}_{x}$. This list is hence
used by the method \emph{computeEuler()} to update the current state using
the standard Euler method. Moreover, the state may be updated by transition
$e$ iff $e$ decrements/increments the considered state component.  At the
end of each step, the method \emph{Norm()} is called to normalize the state
vector taking into account the invariants of the system.

\begin{algorithm}[tbp]
\caption{Algorithm for simulating HSJD systems}
\label{algo1}
\begin{algorithmic}[1]
\Function{SolveSSDE}{$\emph{SSDE},\emph{step},\emph{MaxRuns},\emph{FinalTime}$}
\small
\Statex  $\emph{SSDE}$ = HSJD system.
\Statex  $\emph{step}$ = step used in the Euler solution.
\Statex  $\emph{MaxRuns}$ = maximum number of runs.
\Statex $\emph{FinalTime}$ = maximum time  for each run.
\Statex $\emph{Value}$= vector encoding the state of the HSJD at the current step.
\Statex $\emph{PrValue}$= vector encoding the state at the previous step.
\Statex $\emph{ListESim}$= list of transitions on which discrete simulation step is performed. 
\Statex $\emph{ListE}$= list of transitions on which diffusion step is performed. 
\small
\State \emph{run}=0;
\While {($\emph{run} \leq \emph{MaxRuns})$}
	\State \emph{time} = 0.0;
	\State \emph{SSDE.Init(Value)};\label{line1}
	\While {($\emph{time} \leq \emph{FinalTime})$}
                \State \textbf{print}(\emph{time,Value});
		\State \emph{Value.Copy(PrValue)};\label{line2}
		\State \emph{h} = step;
	    \State \emph{ListESim = SSDE.CheckBound(PrValue)};
		\State \emph{e = ListESim.CheckFire(h)};\label{line3}
		\For{($\emph{SDE} \in \emph{SSDE}$)}
			\State $\emph{ListE=SDE.getEvent(ListESim)}$;
			\State $\emph{SDE.computeEuler(Value,PrValue,h,ListE)}$;
	    	\State \textbf{If} {(\emph{SDE.Check(e)})} \textbf{then} $\emph{SDE.computeSim(Value,PrValue,e)}$;
		\EndFor
		\State \emph{SSDE.Norm(Value)};
		\State \emph{time} += h;
	\EndWhile{}
	\State \emph{run}++;
\EndWhile
\EndFunction
\end{algorithmic}
\end{algorithm}  
   
\label{Algo:1}







\end{document}